\documentclass[fleqn,usenatbib]{mnras}

\usepackage{newtxtext,newtxmath}

\usepackage[T1]{fontenc}

\DeclareRobustCommand{\VAN}[3]{#2}
\let\VANthebibliography\thebibliography
\def\thebibliography{\DeclareRobustCommand{\VAN}[3]{##3}\VANthebibliography}

\usepackage{graphicx}	
\usepackage{amsmath}	
\usepackage{textcomp}	
\usepackage[dvipsnames]{xcolor}
\usepackage{tikz}
\usetikzlibrary{arrows.meta,
                 shapes,
                chains,
                positioning,
                shapes.geometric}
\tikzstyle{arrow}=[draw, -latex]

\newcommand{\appropto}{\mathrel{\vcenter{
  \offinterlineskip\halign{\hfil$##$\cr
    \propto\cr\noalign{\kern2pt}\sim\cr\noalign{\kern-2pt}}}}}

\newcommand{\cs}{c_\mathrm{s}}
\newcommand{\csa}{c_\mathrm{s,adi}}
\newcommand{\csi}{c_\mathrm{s,iso}}

\newcommand{\Mp}{M_\mathrm{p}}
\newcommand{\Mth}{M_\mathrm{th}}
\newcommand{\hp}{h_\mathrm{p}}
\newcommand{\Hp}{H_\mathrm{p}}
\newcommand{\Rp}{R_\mathrm{p}}
\newcommand{\Omegap}{\Omega_\mathrm{p}}
\newcommand{\OmegaK}{\Omega_\mathrm{K}}
\newcommand{\phip}{\phi_\mathrm{p}}
\newcommand{\phipk}{\phi_\mathrm{peak}}
\newcommand{\phil}{\phi_\mathrm{lin}}
\newcommand{\de}{\mathrm{d}}

\newcommand{\pd}[2]{\frac{\partial #1}{\partial #2}}
\newcommand{\td}[2]{\frac{\de #1}{\de #2}}

\newcommand{\vect}[1]{\mathbf{#1}}
\newcommand{\abs}[1]{\left\vert{#1}\right\vert}

\newcommand{\I}{\mathrm{i}}

\newcommand{\RE}{\mathrm{Re}}

\newcommand{\dslin}{\delta \Sigma_\mathrm{lin}}
\newcommand{\lpm}{b_{1/2}^{(m)}} 
\newcommand{\Rcal}{R_\mathrm{cal}} 
\newcommand{\mtr}{m_\mathrm{tr}} 
\newcommand{\mmax}{m_\mathrm{max}} 

\newcommand{\Tex}{\frac{\de {T_\mathrm{ex}}}{\de R}}
\newcommand{\Texs}{\de T_\mathrm{ex}/\de R}
\newcommand{\Textr}{\frac{\de T^\mathrm{tr}_\mathrm{ex}}{\de R}}
\newcommand{\Textrs}{\de T^\mathrm{tr}_\mathrm{ex}/\de R}

\defcitealias{Goodman2001}{GR01}
\defcitealias{Rafikov2002}{Rafikov 2002}
\defcitealias{Cimerman2021}{Paper I}
\defcitealias{GT80}{GT80}

\title[Torque wiggles]{Torque wiggles --- a robust feature of the global disc-planet interaction}

\author[N. P. Cimerman, R. R. Rafikov, and R. Miranda]{
Nicolas P. Cimerman$^{1}$,
Roman R. Rafikov$^{1,2}$\thanks{E-mail: rrr@damtp.cam.ac.uk (RRR)},
Ryan Miranda$^{2}$
\\
$^{1}$Department of Applied Mathematics and Theoretical Physics, University of Cambridge, Wilberforce Road, Cambridge CB3 0WA, UK\\
$^{2}$Institute for Advanced Study, Einstein Drive, Princeton, NJ 08540, USA
}

\date{Accepted XXX. Received YYY; in original form ZZZ}

\pubyear{2022}

\begin{document}

\def\etal{et al.\ \rm}
\def\etal{et al.\ \rm}
\def\Fdw{F_{\rm dw}}
\def\Fdis{F_{\rm dw,dis}}
\def\Fnu{F_\nu}
\def\WD{\rm WD}

\label{firstpage}
\pagerange{\pageref{firstpage}--\pageref{lastpage}}
\maketitle

\begin{abstract}
Gravitational coupling between planets and protoplanetary discs is responsible for many important phenomena such as planet migration and gap formation. The key quantitative characteristics of this coupling is the excitation torque density --- the torque (per unit radius) imparted on the disc by planetary gravity. Recent global simulations and linear calculations found an intricate pattern of low-amplitude, quasi-periodic oscillations in the global radial distribution of torque density in the outer disc, which we call torque wiggles. Here we show that torque wiggles are a robust outcome of global disc-planet interaction and exist despite the variation of disc parameters and thermodynamic assumptions (including $\beta$-cooling).  They result from coupling of the planetary potential to the planet-driven density wave freely propagating in the disc. We developed analytical theory of this phenomenon based on approximate self-similarity of the planet-driven density waves in the outer disc. We used it, together with linear calculations and simulations, to show that (a) the radial periodicity of the wiggles is determined by the global shape of the planet-driven density wave (its wrapping in the disc) and (b) the sharp features in the torque density distribution result from constructive interference of different azimuthal (Fourier) torque contributions at radii where the planetary wake crosses the star-planet line. In the linear regime the torque wiggles represent a weak effect, affecting the total (integrated) torque by only a few per cent. However, their significance should increase in the non-linear regime, when a gap (or a cavity) forms around the perturber's orbit.
\end{abstract}




\section{Introduction}
\label{sec:intro}

Gravitational coupling between a gaseous disc and a perturbing mass (a planet, a satellite, or a binary companion) has been actively explored starting with the seminal studies of \citet{Lin1979} and \citet[][hereafter \citetalias{GT80}]{GT80}. This tidal interaction is recognized as leading to a number of important effects such as gap formation \citep{Lin1986}, planet migration, planetary/binary eccentricity evolution \citepalias{GT80}, disc eccentricity excitation \citep{Lubow1991}, etc. 

A key ingredient of the disc-planet\footnote{In the rest of the paper we will usually call the perturber a 'planet', regardless of its exact nature.} interaction is the excitation of the density waves in the disc --- pressure (sound) waves modified by the differential rotation of the disc. These waves gain energy and angular momentum due to the gravitational potential of the perturber at their launching sites \citepalias[Lindblad resonances,][]{GT80}, propagate through the disc, and eventually deposit their angular momentum and energy to the disc material once they are damped. This makes disc-planet coupling an inherently non-local process \citep{RP12,PR12}. Moreover, the gravity of the perturber acts on the density perturbation due to the density wave resulting in the angular momentum exchange between the perturber and the wave (and eventually the disc).


\begin{figure*}
    \begin{center}
    \includegraphics[width=0.99\textwidth]{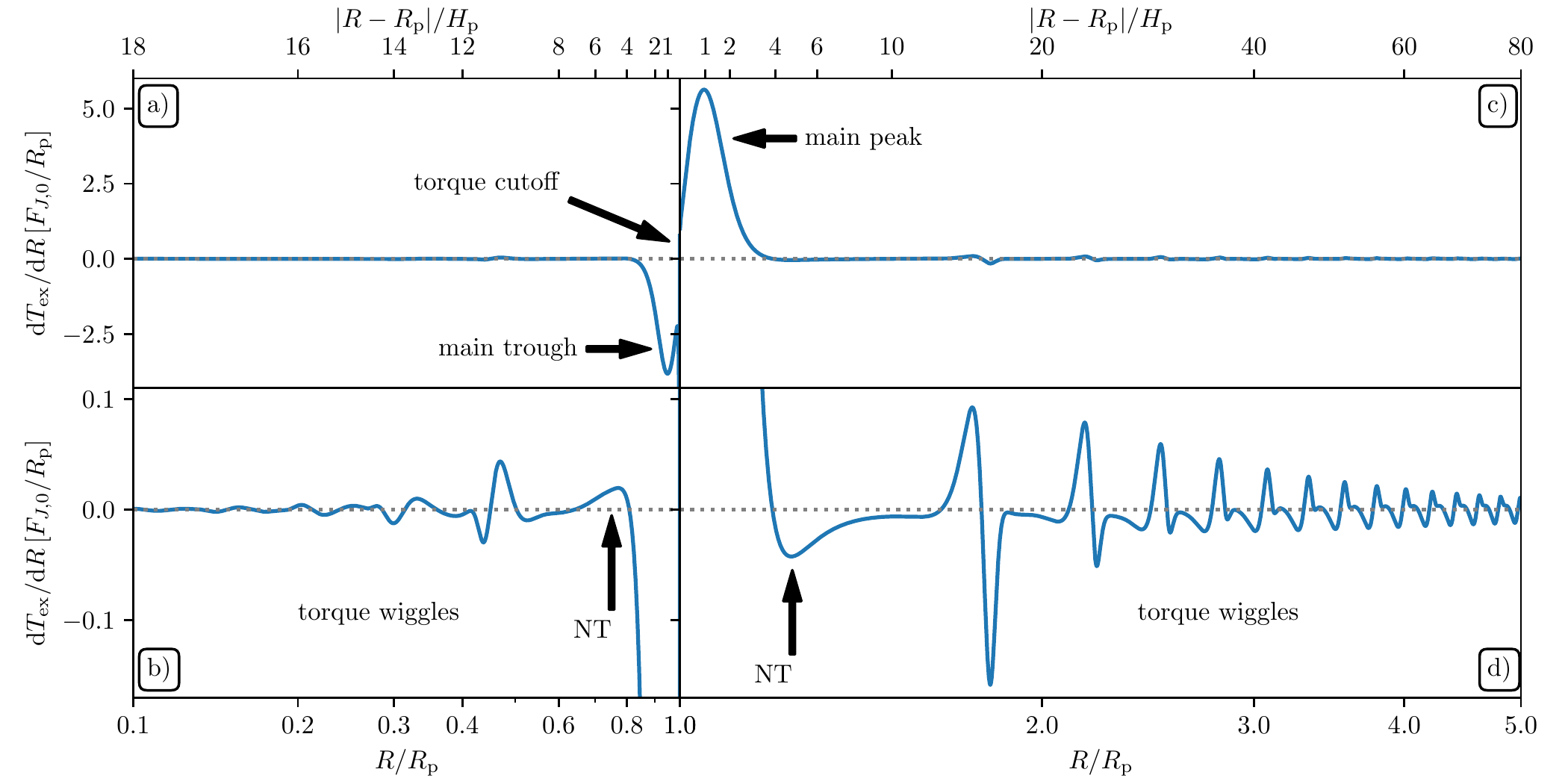}
    \caption{
    Overview of the typical radial distribution of the excitation torque density $\Texs$ (normalized according to the equation (\ref{eq:FJ0})) using linear calculation (Section \ref{sect:calc}) for disc parameters $\hp =0.05$ and $p=q=1$ (se Section \ref{sec:disc_model}). Top and bottom panels show the same data on different vertical scales. The arrows mark the main peaks (`main') of $\Texs$ and indicate the first sign change of $\Texs$ outside the main peak, also known as the {\it negative torque density phenomenon} \citep[`NT', see][]{Dong2011,RP12}. Far away from the planet the label 'torque wiggles' marks the region where $\Texs$ exhibits multiple sign changes \citep{AR18P,Miranda2019I,Miranda2019II,Dempsey2020}.
    }
    \label{fig:dtdr_overview}
    \end{center}
\end{figure*}


One of the main quantitative characteristics of disc-planet coupling is the so-called {\it excitation torque density} $\de T_\mathrm{ex}/\de R$, which is defined in polar cylindrical coordinates $\vect{r}=(R,\phi,z)$ as the $z$-component of the torque exerted by the (Newtonian) potential of the planet per unit radial distance in the disc. One can also interpret $\de T_\mathrm{ex}/\de R$ as the amount of angular momentum added to the density wave per unit radial distance and per unit time by the planetary potential. The integral of $\de T_\mathrm{ex}/\de R$ over the full extent of the disc determines, via Newton's third law, the evolution of the planetary angular momentum, which manifests itself as planetary migration \citepalias{GT80}. The radial profile of $\de T_\mathrm{ex}/\de R$ is one of the key inputs \citep[together with the wave damping mechanism, see][]{GR01,R02} determining the amplitude evolution of the density waves as they travel away from the planetary orbit. Its knowledge is also important for determining the structure of planetary gaps \citep{R02b}. Thus, a detailed understanding of the behaviour of $\de T_\mathrm{ex}/\de R$ is of paramount importance for obtaining a complete picture of disc-planet interaction.

Some of the key features of $\de T_\mathrm{ex}/\de R$ behaviour in a disc which is roughly uniform near the planet (i.e. with no gap formed) have been predicted already in \citetalias{GT80}. In particular, they have shown using linear theory that $\de T_\mathrm{ex}/\de R$ should rapidly decrease as $\vert R-\Rp\vert\lesssim \Hp$ (were $\Rp$ is the semi-major of the planet on circular orbit and $\Hp$ is the disc scaleheight at $\Rp$), an effect that is known as 'torque cutoff'. The overall shape of $\de T_\mathrm{ex}/\de R$ near the planet is set by the tidal disc-planet coupling at numerous Lindblad resonances, with the largest contribution coming from resonances located at $\vert R-\Rp\vert\sim \Hp$. This is indeed what one sees in Fig. \ref{fig:dtdr_overview}, in which we show a typical run of the excitation torque density obtained using our linear calculations, see Section \ref{sect:calc} for details. One can see prominent peak and trough\footnote{The small dent in the trough at $R=\Rp$ is due to the unsaturated corotation torque.} of $\Texs$ located just outside and inside of the planetary orbit, at $\vert R-\Rp\vert\sim \Hp$. The asymmetry in their magnitudes is the reason behind planet migration \citepalias{GT80}. Moreover, both \citet{Lin1979} and \citetalias{GT80} predicted that the excitation torque density should monotonically decay as $\de T_\mathrm{ex}/\de R\propto\vert R-\Rp\vert^{-4}$ for $\vert R-\Rp\vert\gtrsim \Hp$. 

Some of these statements have been refined recently. In particular, the two-dimensional (2D) hydrodynamic simulations of \citet{Dong2011} carried out in the local (sharing sheet) approximation found that $\de T_\mathrm{ex}/\de R$ does not decrease monotonically but actually {\it reverses its sign} around $\vert R-\Rp\vert\approx  3.2\Hp$. This 'negative torque density' phenomenon is illustrated in the inset in Fig. \ref{fig:dtdr_overview}(b), where $\Texs$ first turns negative outside the main positive peak in the outer disc, $R>\Rp$ (in this case it is less pronounced in the inner disc), marked with an arrow (annotated 'NT'). This phenomenon was explained by \citet{RP12} as resulting from interference of the density wave contributions launched at different Lindblad resonances. They have also demonstrated analytically that the decay of $\de T_\mathrm{ex}/\de R$ beyond this point follows $\de T_\mathrm{ex}/\de R\propto \vert R-\Rp\vert^{-4}$ scaling but with a coefficient different in both the sign and magnitude from the prediction of \citetalias{GT80}. This torque reversal was subsequently observed in the global fully nonlinear simulations of \citet{Duffell2012} and \citet{Kley2012}.

Recent global simulations with radially extended domains revealed even more complexity. In particular, 3D simulations by \citet{AR18P} have shown that $\de T_\mathrm{ex}/\de R$ exhibits {\it multiple} sign reversals at $\vert R-\Rp\vert\sim \Rp$ (see their Fig. 3). Evidence for this behaviour can also be found in the linear  calculations of \citet[][see bottom rows of their Figs. 7 \& 8]{Miranda2019I} and simulations of \citet[][see their Fig. 1]{Miranda2019II}, visible as the oscillations of the density wave angular momentum flux $F_J$(its radial derivative is equal to $\de T_\mathrm{ex}/\de R$ in the absence of wave damping), which are especially pronounced in the outer disc. Sign reversals of $\de T_\mathrm{ex}/\de R$ in the outer disc can also be seen in the results of \citet[][see the top row of their Fig. 5]{Dempsey2020} obtained for low perturber masses, although their simulations also included viscosity and formation of a gap around the planet. Our Fig. \ref{fig:dtdr_overview}b,d also shows such $\Texs$ features very clearly both as multiple, rather regularly spaced wiggles in the outer disc for $R\gtrsim 2\Rp$ as well as the lower-amplitude, irregular features in the inner disc.

The goal of our present work is to shed light on the origin of the sign reversals and oscillations of $\de T_\mathrm{ex}/\de R$. In particular, we demonstrate that these features are indeed real and not a numerical artefact. It is true that in all aforementioned cases, which fall in the linear regime of tidal coupling, the amplitude of these features is rather small, not exceeding several per cent of the main peak of the torque density (and decreasing with the distance from the planet). However, the magnitude of $\de T_\mathrm{ex}/\de R$ oscillations grows as the mass of the perturber increases and the disc-perturber coupling becomes nonlinear, see \citet{Dempsey2020}.  Thus, clarifying the origin of the features of $\de T_\mathrm{ex}/\de R$ behaviour in the linear regime will also help us understand the torque density behaviour in the nonlinear regime, relevant for circumbinary discs around stellar and supermassive black hole binaries (Cimerman \& Rafikov, in prep.). 

Our work is organized as follows. After describing our setup and methods in Section \ref{sec:math}, we provide a heuristic explanation for the torque wiggles in Section \ref{sec:heuristic}. We then present a detailed theoretical model for the origin and properties of the torque wiggles in Section \ref{sec:theory}, which may be skipped at first reading, and explore their sensitivity to the disc parameters in Section \ref{sec:var_par} and to thermodynamic assumptions in Section \ref{sec:res_beta}. We further discuss our results in Section \ref{sec:disc} and briefly summarize them in Section \ref{sec:disc_conclusion}.


\section{Setup and methods}
\label{sec:math}



\subsection{Model setup}
\label{sec:disc_model}


We consider a 2D model of a (initially axisymmetric) razor-thin disc that orbits a central star of mass $M_\star$. We adopt polar $(R,\phi)$ coordinates centred on the central star. The disc has aspect ratio $h \equiv H/R \ll 1$, where $H = \cs/\Omega$ is the vertical scale-height, $\cs$ is the sound speed and $\Omega$ is the angular orbital frequency. We assume that the effective viscosity of the disc gas is low, implying that the flow is laminar (not turbulent), and do not include any explicit viscosity in our model. Nor do we include the disc self-gravity. 

This disc is perturbed by the gravitational field of a coplanar planet with mass $\Mp$, moving on a circular orbit with radius $\Rp$, giving rise to non-axisymmetric perturbations to the basic state. Although our coordinate frame is centred on a (moving) central star, for simplicity we do not include the indirect potential when computing the response of the disc due to the presence of the planet. In this regard, we follow a large number of existing studies of disc-planet coupling that also neglected the indirect potential \citep{Bate2003,DA2008,DA2010,Duffell2012,Dong2011,Dong2011b,RP12,Miranda2019I,Miranda2019II,Miranda2020I,Fairbairn2022}. We note that the studies fully accounting for the indirect potential of the planet \citep{Kley2012,AR18P} did not find noticeable differences from the calculations neglecting it. We also do not allow for the orbit of the planet to evolve with time.

In this work, we limit ourselves to planet masses that are well below the {\it thermal mass}
\begin{align}
	\Mth=\frac{c_\mathrm{p}^3}{\Omega_\mathrm{p} G}=\left(\frac{\Hp}{\Rp}\right)^3 M_\star=\hp^3 \, M_\star,
	\label{eq:Mth}
\end{align}
where $c_\mathrm{p} = \cs(\Rp)$ is the sound speed at $\Rp$, $\Omega_\mathrm{p}$ is the orbital angular frequency at $\Rp$, $\Hp = H(\Rp)$ and $\hp=h(\Rp)$. The assumption $\Mp\lesssim\Mth$ implies that the disc response to the planetary gravity is linear, i.e. $\delta \Sigma / \Sigma \sim \Mp /\Mth \ll 1$, where $\delta \Sigma = \Sigma - \Sigma_0$ is the planet-induced perturbation of the surface density $\Sigma$ relative to the background surface density $\Sigma_0(R)$. 

Regarding the structure of the unperturbed (by the planet) disc, we assume that surface density and temperature obey a power-law ansatz
\begin{align}
    \Sigma_0(R) & = \Sigma_0(\Rp) \left(\frac{R}{\Rp}\right)^{-p},
    \label{eq:Sig_i}\\
   T_0(R) & = T_0(\Rp) \left(\frac{R}{\Rp}\right)^{-q},
    \label{eq:T_i}
\end{align}
such that the isothermal sound speed in the disc is
\begin{align}
    \csi(R) = \hp \Omegap \Rp  \left(\frac{R}{\Rp}\right)^{-q/2}.
    \label{eq:csi_i}
\end{align}
The disc is initially in radial centrifugal balance, taking into account the radial pressure gradient:
\begin{align}
    \Omega^2 &= \OmegaK^2 + \frac{1}{R \Sigma} \td{P}{R}
    = \OmegaK^2 \left[1 - h^2(R)(q+p) \right],
    \label{eq:centr}
\end{align}
where $\OmegaK^2 = \sqrt{GM_\star/R^3}$ is the Keplerian orbital frequency, and $P$ is the gas pressure (see below). The unperturbed velocity of gas is given by $u_{R,0} = 0$, $u_{\phi,0}(R) = R \Omega_0(R)$, with $\Omega_0(R)$ defined by equation (\ref{eq:centr}) using $\Sigma_0, P_0$.

Regarding the equation of state (EoS), we explore several options. For most of our calculations we adopt the adiabatic relation 
\begin{align}
    P = K\Sigma^{\gamma},
    \label{eq:adi}
\end{align}
where we fix the adiabatic exponent $\gamma = 7/5$ and $K$ is the adiabatic constant. In our simulations we explicitly solve the energy equation, see Appendix \ref{sec:num_methods}, thus properly accounting for the evolution of $K$ (which would be conserved for each fluid element in a Lagrangian sense only in the absence of shocks or viscosity). It is important to keep in mind that the equation (\ref{eq:T_i}) does not imply the locally-isothermal EoS: it is used, together with the equation (\ref{eq:Sig_i}), to set the initial radial profile of $K$ in our adiabatic calculations. The adiabatic sound speed is $\csa = \gamma^{1/2} \csi$.

Additionally, in Section \ref{sec:res_beta} we consider discs that allow for thermal relaxation of perturbations towards the unperturbed background following the so-called $\beta$-cooling prescription. We describe its detailed implementation in Appendix \ref{sec:beta}.

The treatment of thermodynamics has important implications on the global propagation of density waves excited by a planet. In particular, thermal physics directly affects the evolution of the angular momentum flux (AMF) associated with these waves, defined as
\begin{align}
    F_J(R) &= R^2 \oint \Sigma(R,\phi) u_R(R,\phi) \delta u_\phi(R,\phi) \,\de \phi,
    \label{eq:FJ}
\end{align}
with velocity perturbations $u_R$ and $\delta u_\phi(R,\phi)=u_\phi(R,\phi)-u_{\phi,0}(R)$. \citet{Miranda2020I,Miranda2020II} have shown that for adiabatic discs obeying (\ref{eq:adi}) the AMF carried by the density wave is conserved in the absence of wave damping (see also Section \ref{sec:planet_waves}). On the contrary, discs with thermal relaxation, including the locally isothermal discs \citep{Miranda2019II}, do not conserve the wave AMF, which may decay or get amplified even in the absence of non-linear damping. To distinguish these possibilities, we will call our models {\it AMF-preserving} when the disc thermodynamics is such (i.e. adiabatic) that the wave AMF is conserved in the absence of explicit damping, linear or nonlinear.

Linear calculations of the integrated one-sided Lindblad torque \citepalias{GT80} find the characteristic magnitude of the AMF of the planet-driven density waves to be
\begin{align}
    F_{J,0} = \left(\frac{\Mp}{M_\star}\right)^2 \hp^{-3} \Sigma_\mathrm{p} \Rp^4 \Omegap^2,
    \label{eq:FJ0}
\end{align}
which we will use as a reference value for $F_J$. To allow for meaningful comparison, radial torque densities are given in units $F_{J,0}/\Rp$, as in Fig. \ref{fig:dtdr_overview}.


\subsection{Methods and torque calculation}
\label{sect:calc}


We employ two methods to calculate the structure of a disc perturbed by a planet. First, we determine the disc structure in the linear approximation valid in the limit $\Mp/\Mth\ll 1$. The method for solving the linear problem has been developed in \citet{Miranda2019I,Miranda2020I} and we employ it here as well. Some details of its implementation and the parameters used can be found in Appendix \ref{sec:lin_solv}. 

Second, we also perform fully non-linear simulations (see Appendix \ref{sec:setup_nl} for details and numerical parameters) of the problem using Athena++\footnote{Athena++ is publicly available on \href{https://github.com/PrincetonUniversity/athena/}{GitHub}.} \citep{Athenapp2020}, even though we still use $\Mp=0.01\Mth$ in these runs, which is well in the linear regime. This allows us to cross-check our linear calculations. 

The perturbed pattern of the disc surface density $\Sigma(R,\phi)$ obtained by either of these two methods is then used to verify the results of our semi-analytical calculations of the torque wiggles in Section \ref{sec:theory}. More specifically, we compute the excitation torque density (torque per unit radius $\de R$) by integrating ${\bf R}\times \de {\bf F}/\de R$ (with $\de {\bf F}$ being the direct gravitational force exerted by a planet onto a disc element $\de S=R \de R \de \phi$) over $\phi$ at fixed $R$ as
\begin{align}
\frac{\de \vect{T}_\mathrm{ex}}{\de R}
 =G \Mp\int\limits_0^{2\pi}\frac{{\bf R}\times ({\bf \Rp}-{\bf R})}{|{\bf R}-{\bf \Rp}|^3}\Sigma(R,\phi) R \,\de \phi,
\label{eq:dTvect}
\end{align}
which points in the $z$-direction. Note that this calculation accounts only for the {\it direct} planetary potential. In other words, we do not include the {\it indirect} potential in the torque calculation, to be consistent with all existing studies. Equation (\ref{eq:dTvect}) uses a purely Newtonian potential for torque calculation, which is somewhat different from the softened potential (\ref{eq:phi4_p2}) used in our simulations. However, the two are essentially the same at the separations of interest for us.

We perform a parameter study of the problem by varying
disc properties: surface density and temperature slopes $p$ and $q$ and aspect ratio at the planet radius $\hp$. In Table \ref{tab:param} we give an overview of the disc parameter sets that are explored in this work. We indicate the number of modes $\mmax$ obtained for the solution of the linear problem (see Appendix \ref{sec:lin_solv}) and mark the parameter sets for which the Athena++ simulations were performed. The adiabatic model with $\hp=0.1$ and $p=q=1$ highlighted in bold in Table \ref{tab:param} is the {\it fiducial} disc model that we will use to illustrate many of our results in Sections \ref{sec:heuristic} \& \ref{sec:theory}. 


\begin{table}
\begin{tabular}{lcccccr}
 \hline
 $\hp$ & $p$ & $q$ & EoS & $\beta$ & $\mmax$ & Athena++ sim?\\
 \hline
 {\bf 0.1 } & {\bf 1 } & {\bf 1 } & {\bf AD } & {\bf - } & {\bf 100 } & {\bf \checkmark }\\
 0.1 & 0 & 1 & AD & - & 100 &\checkmark \\
 0.1 & 1 & 1/2 & AD & - & 100 & \checkmark \\
 0.1 & 1 & 0 & AD & - & 100 & \checkmark \\
 0.05 & 1 & 1 & AD & - & 220 & \checkmark \\
 0.025 & 1 & 1 & AD & - & 320 & $\times$ \\
 0.1 & 1 & 1 & TR & $10^2$ & 100 & \checkmark \\
 0.1 & 1 & 1 & TR & $10$ & 100 & \checkmark \\
 0.1 & 1 & 1 & TR & $1$ & 100 & \checkmark \\
 0.1 & 1 & 1 & TR & $10^{-2}$ & 100 & \checkmark \\
 0.1 & 1 & 1 & TR & $10^{-3}$ & 100 & \checkmark \\
 0.1 & 1 & 1 & TR & $10^{-4}$ & 100 & \checkmark \\
 0.1 & 1 & 1 & TR $\rightarrow$ LI & $10^{-6}$ & 100 & \checkmark \\
 \hline
\end{tabular}
\caption{
Sets of parameters considered. From left to right, the first
five columns give the scale-height at the planets radial location $\hp$, surface density slope $p$, temperature slope $q$, the equation of state (adiabatic [AD], thermal relaxation [TR] or locally isothermal [LI)]. For the adiabatic cases it is implied that linear models use $\beta = 10^2$ and cooling is entirely switched off in Athena++ simulations. The two rightmost columns indicate the maximum mode number $\mmax$ (Appendix \ref{sec:lin_solv}), and whether a non-linear Athena++ simulation was performed. Values in boldface correspond to the fiducial disc model.
}
\label{tab:param}
\end{table}

\begin{figure}
    \begin{center}
    \includegraphics[width=0.49\textwidth]{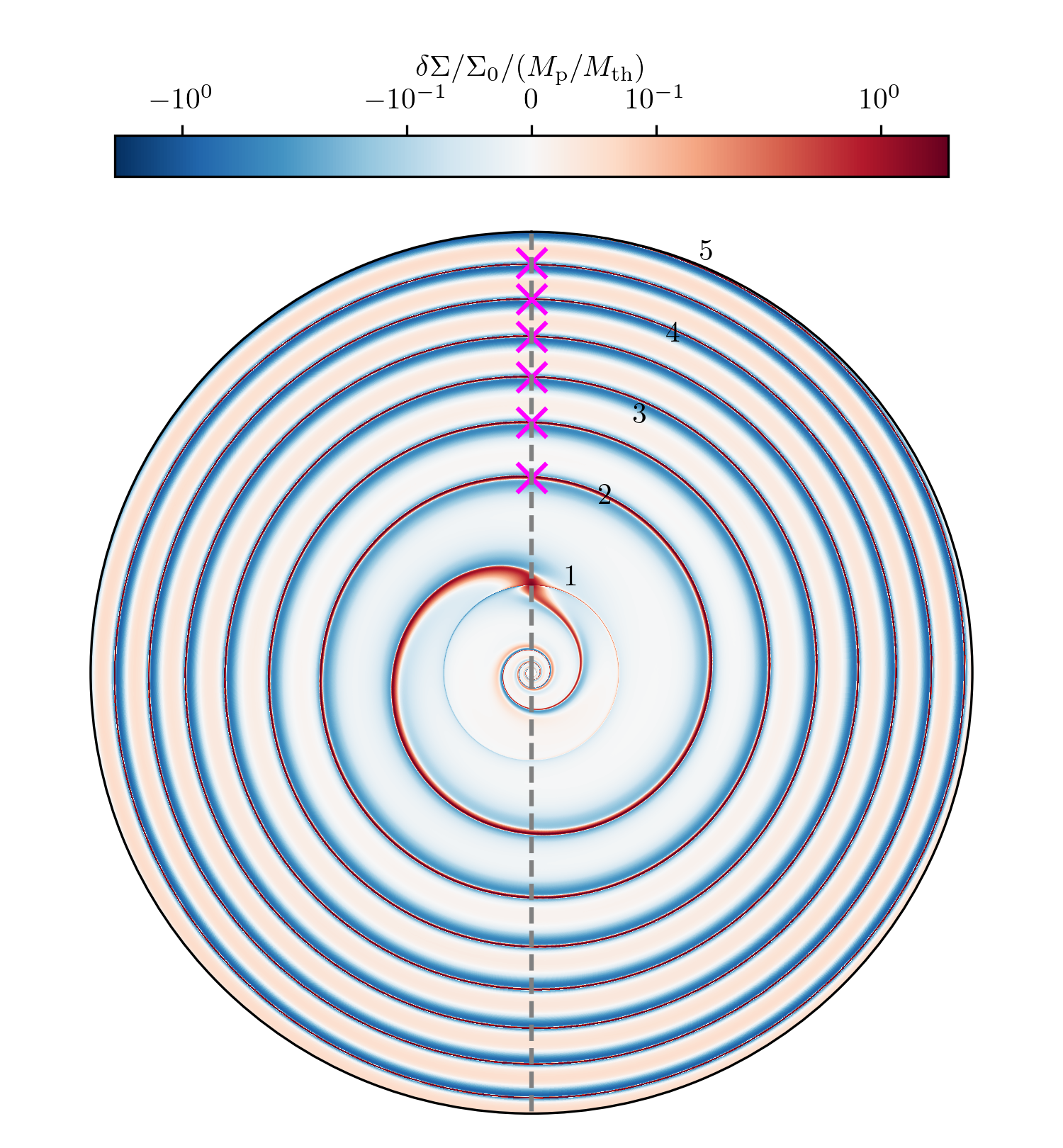}
    \vspace*{-0.5cm}
    \caption{
        Polar plot of $\delta \Sigma/\Sigma_0 \times (\Mp/\Mth)^{-1}$ for the fiducial disc model. The grey dashed line
        goes through the origin where the central star is located and the planet. Pink crosses mark the locations where the peak of the spiral density wave $\phi=\phipk(R)$ crosses the line $\phi = \phip$; we only show them in the outer disc for clarity. The planet and disc rotate in a clockwise direction.
    }
    \label{fig:polar_dsigma}
    \end{center}
\end{figure}

\begin{figure*}
    \begin{center}
    \includegraphics[width=0.99\textwidth]{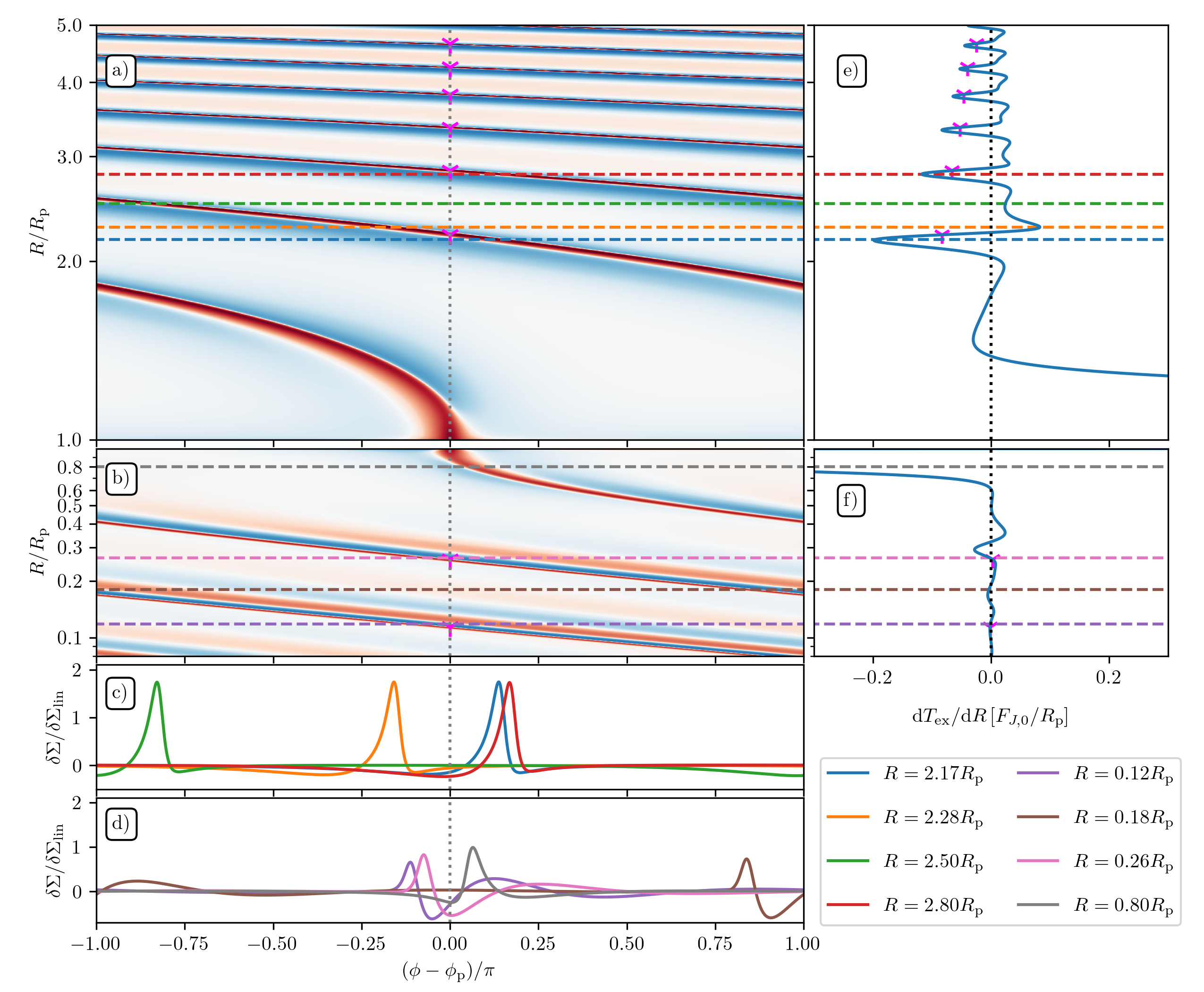}
    \vspace*{-0.2cm}
    \caption{(a),(b) 2D Map of the surface density perturbation obtained by linear solution in the fiducial disc model ($\hp=0.1$, $p=q=1$) together with (c),(d) azimuthal slices of the surface density perturbation at several radii, and (e),(f) the radial profile of torque density, in the inner and outer disc respectively. The colormap is the same as in Fig. \ref{fig:polar_dsigma}. Azimuthal slices in panels (c) and (d) correspond to the radii indicated by the horizontal coloured dashed lines (of corresponding color) in panels (a),(b),(e) and (f). Pink tripods indicate locations where the peak of the wake crosses the line connecting the central star and the planet ($\phipk = \phip$) as in Fig. \ref{fig:polar_dsigma}; here we also show them in the inner disc. See text for details.
    }
    \label{fig:tq_map}
    \end{center}
\end{figure*}
    


\section{Heuristic explanation of the torque wiggles}
\label{sec:heuristic}


In this section we provide a simple heuristic explanation for the origin of the torque wiggles seen in Fig. \ref{fig:dtdr_overview}, and for the difference in their appearance in the outer and inner parts of the disc. This argument is then buttressed with quantitative calculations in Section \ref{sec:theory}.

In Fig. \ref{fig:polar_dsigma} we display a polar 2D map of the global surface density perturbation obtained via a linear calculation for the fiducial disc model: adiabatic with $\hp=0.1$ (higher than $\hp=0.05$ used in Fig. \ref{fig:dtdr_overview}) and $p=q=1$. One can easily see a one-armed spiral pattern that the planet excites inside and outside its orbit. To zeroth order, the shape of this spiral is given by the curve ($R, \phil(R)$), where\footnote{This result assumes $\cs=\csa$ in AMF-preserving discs and $\cs=\csi$ in isothermal discs. Things get more complicated in discs with cooling, see \citet{Miranda2020I}.} \citep{R02,OL02}
\begin{align}
\phi_\mathrm{lin} & = \phi_\mathrm{p} + \varphi(R),
\label{eq:phi-lin}\\
\varphi(R) &=\mathrm{sign}(R-\Rp) \int\limits_{R_\mathrm{p}}^R \frac{\Omega(R^\prime)-\Omegap}{c_\mathrm{s}(R^\prime)} \de R^\prime 
\label{eq:Deltaphi}
\end{align}
For a Keplerian disc ($\Omega \rightarrow \OmegaK$) and sound speed profile in the form (\ref{eq:csi_i}) one finds
\begin{align}
\varphi(R) & = \mathrm{sign}(R-\Rp) \hp^{-1}\left[g\left(\frac{R}{\Rp}\right) - l\left(\frac{R}{\Rp}\right)\right],
\label{eq:Deltaphi2}\\
g(x) & =
\begin{cases}
\frac{2}{q-1}\left[x^{(q-1)/2} - 1\right] & (q \neq 1), \\
\ln(x) & (q = 1),
\end{cases}
\label{eq:gx}\\
l(x) & = \frac{2}{q+2}\left[x^{(q+2)/2}-1\right].
\label{eq:lx}
\end{align}

This approximation works very well in the outer disc, as shown in \citet{Miranda2019I}. We see that over the entire domain $R>\Rp$ the spiral maintains a sharp crest (red), neighboured by a lower amplitude trough (blue). But in the inner disc this approximation starts failing for $R\lesssim 0.6\Rp$, as the spiral arm splits into multiple peaks as it propagates. This evolution of a single-armed pattern into multiple spiral arms was found in simulations \citep[e.g.][]{Dong2015,Fung2015} and explained by \citet{BZ18a} and \citet{Miranda2019I}. This difference\footnote{\citet{Miranda2019I} demonstrated that the formation of a secondary spiral arms is also possible in the outer disc but only for small values of $\hp$. We start to see some evidence for this for $\hp = 0.025$, see Section \ref{sec:wake_pars}.} has important implications for the shape of torque wiggles in different regions of the disc. 

In Fig. \ref{fig:polar_dsigma} we also show the dashed grey line that passes through both the star and planet, i.e. $\phi = \phip = $ const. As it turns out, this line has a special significance for explaining the nature of the torque wiggles. The crosses (not shown in the inner disc to avoid confusion) indicate the locations where the planet wake crosses this line, i.e. $\phipk = \phip$ modulo $2\pi$. We will refer to this situation as {\it wave-} or {\it wake-crossings}. In the outer disc, where the approximation (\ref{eq:phi-lin})-(\ref{eq:lx}) works well, the radius $R_n$ of $n$-th such crossing ($n=0,1,2,\ldots$) is given by the condition $\varphi(R_n)=2\pi n$, with $R_0=\Rp$.

To further support our statements, in Fig. \ref{fig:tq_map} we show the same 2D map of the relative surface density perturbation $\delta \Sigma / \Sigma_0$ as in Fig. \ref{fig:polar_dsigma} but now in Cartesian geometry, see panels (a) and (b) for inner and outer disc, respectively. Also, in panels (c) \& (d) we show the azimuthal profiles of $\delta \Sigma$ in the outer and inner disc, respectively, at several fixed radii (shown in panels (a) \& (b) using dashed lines of the same color). They are normalized by the characteristic wave amplitude $\dslin(R)$ expected in the linear regime (see Section \ref{sec:planet_waves}) to highlight the evolution of the shape of the density wake. Finally, in panels (e) \& (f) we show the radial profile of the excitation torque density $\Texs$ on the same radial interval as in panels (a) and (b) to allow cross-matching of different features. We now examine this figure separately for the outer and inner discs.


\subsection{Outer disc}
\label{sec:outer_typical}


In the outer disc ($R>\Rp$) the planet-driven spiral retains a narrow one-armed shape as it wraps around the full azimuthal range of the disc multiple times. Moreover, panel (c) shows that the azimuthal profile of $\delta \Sigma$ maintains its shape (upon normalization by $\dslin$) to a good accuracy as the wake propagates, suggestive of a self-similar evolution. At the same time, the azimuthal location of the wake with respect to the $\phi=\phip$ line (vertical dotted line) steadily changes, affecting the sign and value of $\Texs$.

At $R = 2.17 \Rp$ (blue lines), the wake approaches its first passing of $\phi = \phip$ and the torque density displays a sharp minimum. Panel (c) reveals that at this radius $\delta \Sigma$ (blue curve) has a (positive) peak at $\phi> \phip$ and (negative) trough at $\phi< \phip$, with the transition happening very close to $\phi = \phip$, when the wake is closest to the planet.
In this optimal situation both the wake overdensity (relative to unperturbed disc) at $\phi> \phip$ and the wake underdensity at $\phi< \phip$ pull the planet forward by their gravity, increasing its angular momentum. Newton's third law then implies that the wake must be losing its angular momentum at this radius, resulting in strongly negative\footnote{Even despite the smaller, leading trough at $\phi - \phip \simeq 0.2\pi$, which provides a small but positive contribution to $\Texs$.} $\Texs$, which is what we see in panel (e).

As the wake crosses the line $\phi = \phip$ (marked with a tripod), the wake in panel (c) shifts to $\phi<\phip$, pulling back on the planet, which gives rise to a positive peak of $\Texs$ at $R=2.28\Rp$ (orange curve). This maximum is roughly only half of the magnitude of the preceding minimum, which is caused by the fact that the wake shape is not perfectly symmetric in $\phi$ (otherwise we would expect similar magnitude of neighbouring extrema for tightly wound waves). Moving further out to $R=2.5\Rp$ (green curves), the peak of the wake is close to $\phi=\phip + \pi$ (i.e. behind the star relative to the planet), which lowers (still positive) $\Texs$. 

Beyond that point $\Texs$ becomes negative again and at $R \simeq 2.8 \Rp$ (red lines) the wake profile in panel (c) almost coincides azimuthally with that at the previous minimum of $\Texs$ (blue curve). As a result, $\Texs$ reaches another (negative) minimum, although not as deep as the first one since this time the wake is further from the planet.

This pattern of alternating maxima and minima of $\Texs$ (with steadily decaying amplitude) repeats in a very regular way. Comparing panels (a) and (e), it is clear that the radial periodicity of this pattern in dictated by the repeated wake crossings of the $\phi = \phip$ line as the spiral propagates, with sharp minima occurring just before azimuthal alignments of the surface density peak with the planet. The overall shape of $\Texs$ is only quasi-periodic: closer to the planet $\Texs$ displays two local maxima (as wakes fully wraps around the star), while further out in the disc there is only one maximum. Nevertheless, the radial periodicity of this slowly evolving pattern of $\Texs$ is very accurately set by the wake crossings, thanks to the essentially self-similar shape of $\delta\Sigma$ in the outer disc.


\subsection{Inner disc}
\label{sec:inner_typical}


We now turn to the inner disc ($R < \Rp$), see panels (b), (d), (f). Close to the planet, at $\vert R-\Rp \vert \simeq (2-3)\Hp = (0.2-0.3) \Rp$, the azimuthal profile of $\delta \Sigma$ is similar to that in the outer disc, except that now it propagates in the opposite direction relative to the local mean flow. This can be seen in the grey curve in panel (d) drawn for $R=0.8\Rp$ (sampling the main trough of $\Texs$), which is similar in shape (upon $\phi$-reflection and rescaling) to the curves in panel (c). 
 
However, closer to the star, and different from the wake in the outer disc, we notice an additional peak (red) in the surface density perturbation in panel (b) appearing next to the trough (blue) at $R\lesssim 0.6\Rp$. In the innermost regions, this secondary spiral arm is well-developed and there is a hint of a tertiary arm (peak) forming. This picture of multiple arm formation is consistent with \citet{Bae17} and \citet{Miranda2019I}. These modifications are also reflected in the $\delta\Sigma/\dslin$ profiles for $R = 0.265 \Rp, 0.18 \Rp,  0.118 \Rp$ in panel (d), which develop multiple peaks and troughs and certainly do not evolve in a self-similar fashion like in the outer disc (see also Figs. \ref{fig:psi_var_pq},\ref{fig:psi_var_h}). 

The more complicated and steadily evolving (in $R$) azimuthal profiles of $\delta\Sigma$ lead to the loss of coherence of $\Texs$ at consecutive wake crossings in the inner disc, e.g. at $R = 0.265 \Rp$ (pink) and $R= 0.118 \Rp$ (magenta). As a result, one no longer sees the radial quasi-periodicity of the $\Texs$ features, which are so obvious in the outer disc. Moreover, the amplitude of these features is also significantly reduced, partly because of the lower angular momentum flux carried by the wake in the inner disc (compare the magnitudes of the main peak and trough near the planet in Fig. \ref{fig:dtdr_overview}) but also because of faster radial decay of the torque density in the inner disc, see Sections \ref{sec:theory} \& \ref{sec:asy_analysis}.  

\begin{figure}
    \begin{center}
    \includegraphics[width=0.49\textwidth]{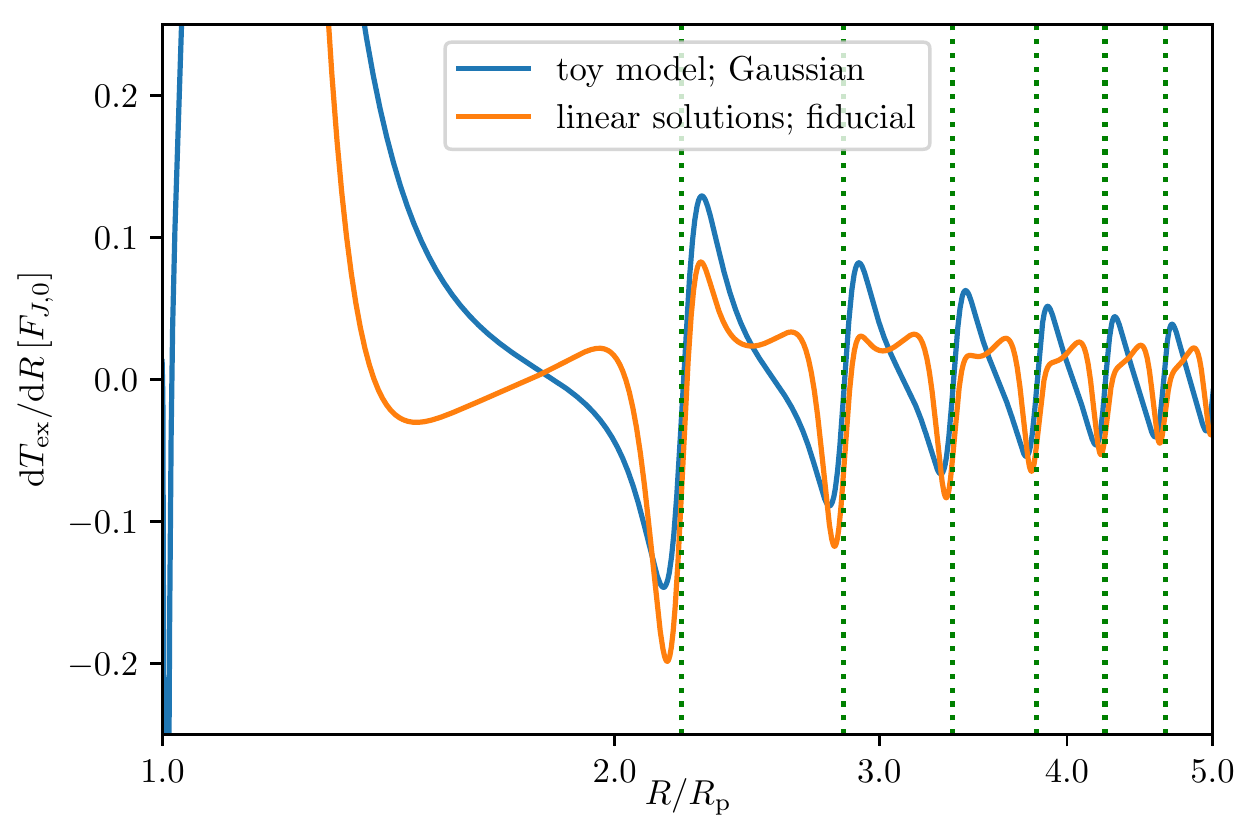}
    \vspace*{-2em}
    \caption{
        Excitation torque density in the outer disc for our toy Gaussian model (blue) and the full linear solution of the linear problem (orange). Note that they share the same periodicity, confirming that it is the behaviour of $\phil(R)$ that determines the radial structure. The toy model produces a more symmetric pattern with neighbouring minima and maxima being of almost identical height. The torque density associated with the Gaussian profile vanishes exactly when $\phipk = \phil$,
        as expected from symmetry.
    }
    \label{fig:toy_dtdr}
    \end{center}
\end{figure}


\subsection{Toy model of the wiggles}
\label{sec:toy}


The observations made in Sections \ref{sec:outer_typical} and \ref{sec:inner_typical} strongly suggest that the approximate self-similarity of the wake shape (along the azimuthal direction) as $R$ varies is the critical factor for developing a regular pattern of the torque wiggles. This self-similarity exists in the outer but not in the inner disc, resulting in a rather irregular pattern of the wiggles for $R<\Rp$.

To emphasize the role of self-similarity of $\delta\Sigma$ even further, in Fig. \ref{fig:toy_dtdr} we show $\Texs$ computed in the outer disc for a hypothetical 'Gaussian' wake in the form
\begin{align}
    \delta \Sigma (R, \phi) = A \, \dslin(R) \,\exp\left[
    -\frac{1}{2} \left( \frac{\phi-\phil(R)}{w} \right)^2
    \right],
    \label{eq:Gauss}
\end{align}
where $A$ is an (arbitrary) amplitude and we set $w=\hp$ as the characteristic azimuthal width of the wake. This wake has a correct radial scaling of its amplitude to ensure the conservation of the wave angular momentum flux (see Section \ref{sec:planet_waves}) and follows the curve $\phi=\phil(R)$ in polar coordinates. 

One can see that this artificial wake gives rise to a regular pattern of torque wiggles (blue) with a clear radial periodicity and decaying amplitude. Because of the symmetric shape of the Gaussian wake, the consecutive peaks and troughs of the wiggles have similar amplitude as compared to the actual planet-driven $\Texs$ (orange) also shown in that figure, see Section \ref{sec:outer_typical}. But the pattern of radial variability is the same for both curves, highlighting that it is the $\phil(R)$ behaviour (setting the radial frequency of wake crossings) that determines it.


\section{Theoretical model of the torque wiggles}
\label{sec:theory}


We now provide a more refined mathematical description of the torque wiggles. Our goal is to provide a semi-analytical description of these features in the outer disc, that would allow one to reproduce not only their radial (quasi-)periodicity but also their amplitude. We will demonstrate that this is possible in some cases of physical importance.  

We start by introducing the planet-induced perturbation of the surface density $\delta\Sigma(R,\phi)=\Sigma(R,\phi)-\Sigma_0(R)$. Since $\Sigma_0$ is axisymmetric and gives no net contribution to the torque, we can replace $\Sigma(R,\phi)$ with $\delta\Sigma(R,\phi)$ in equation (\ref{eq:dTvect}). Always working in the frame co-rotating with the planet such that $\phip = 0$, we can then write the magnitude of the torque density in the outer disc as
\begin{align}
\Tex &= G \Mp R^2 \Rp \int_0^{2\pi}\frac{\sin\phi~\delta\Sigma(R,\phi) }{\left(R^2+\Rp^2-2R \Rp\cos\phi\right)^{3/2}} \,\de \phi
\nonumber\\
 &= G \Mp \,\alpha\int_0^{2\pi}\frac{\sin\phi~\delta\Sigma(R,\phi)}{\left(1+\alpha^2-2\alpha\cos\phi\right)^{3/2}} \,\de \phi,
\label{eq:dT}
\end{align}
where we introduced $\alpha \equiv \Rp/R<1$.

Integrating equation (\ref{eq:dT}) by parts we find
\begin{align}
\Tex = -G \Mp \int_0^{2\pi}\frac{\partial\delta\Sigma/\partial\phi}{\left(1+\alpha^2-2\alpha\cos\phi\right)^{1/2}} \de \phi.
\label{eq:dT1}
\end{align}
Using the series expansion
\begin{align}
\frac{1}{\left(1+\alpha^2-2\alpha\cos\phi\right)^{1/2}} = \frac{1}{2}b_{1/2}^{(0)}(\alpha)
+\sum\limits_{j=1}^\infty b_{1/2}^{(j)}(\alpha)\cos (j\phi),
\end{align}
where
\begin{align}
    b^{(j)}_{1/2} (\alpha) = \frac{1}{\pi}
                    \int_{0}^{2\pi}
                        \frac{\cos (j \phi)\,\de \phi}{(1 + \alpha^2 - 2 \alpha \cos \phi)^{1/2}},
    \label{eq:lap_def}
\end{align}
are the Laplace coefficients \citep[e.g.][]{MurrayDermott},
one can re-write equation (\ref{eq:dT}) as
\footnote{The term with $b_{1/2}^{(0)}(\alpha)$ vanishes since it is $\propto \int_0^{2\pi} \partial_\phi \delta \Sigma \, \de \phi = 0$.}
\begin{align}
\Tex &= -GM_p \sum \limits_{j=1}^\infty b_{1/2}^{(j)}(\alpha)\int_0^{2\pi}\frac{\partial\delta\Sigma}{\partial\phi}\cos (j\phi) \,\de\phi.
\label{eq:dT2}
\end{align}
Up to a normalization, we recognize the remaining integral as the cosine Fourier coefficient of $\partial\delta\Sigma/\partial\phi$:
\begin{align}
C_j^c(R) = \frac{1}{\pi} \int_0^{2\pi}\frac{\partial\delta\Sigma}{\partial\phi}\cos (j\phi) \,\de \phi,
\label{eq:Cj}
\end{align}
such that we can write equation (\ref{eq:dT2}) as
\begin{align}
\Tex = -\pi G \Mp \sum \limits_{j=1}^\infty b_{1/2}^{(j)}(\alpha)C_j^c(R),~~~~~\alpha=\frac{\Rp}{R}<1,
\label{eq:dT3}
\end{align}
where all information about the surface density perturbation is contained in the $C_j^c(R)$. 

In the inner disc, defining $\alpha\equiv R/\Rp<1$, an analogous calculation gives
\begin{align}
\Tex = -\pi G \Mp \,\alpha \sum \limits_{j=1}^\infty b_{1/2}^{(j)}(\alpha)C_j^c(R),~~~~~\alpha=\frac{R}{\Rp}<1,
\label{eq:dT3_in}
\end{align}
with an extra factor of $\alpha$ compared to the equation (\ref{eq:dT3}). This has important consequences for the amplitude of the torque wiggles in the inner disc, see Section \ref{sec:asy_analysis}.

So far we have not assumed any particular form of $\delta \Sigma$ (and $\partial_\phi \delta\Sigma$) such that our results (\ref{eq:dT2}) \& (\ref{eq:dT3}) are fully general.


\subsection{Torque wiggles for self-similar spiral arms}
\label{sec:selfim}


In the rest of the paper, unless mentioned otherwise, we will focus on exploring the properties of the torque wiggles in the outer disc. Inspired by our observations in Section \ref{sec:outer_typical} that, once formed, the outer spiral arm maintains its shape without significant distortion (see Fig. \ref{fig:tq_map}(c)), we provide a more in-depth characterization of the torque wiggles in the case of a fully \textit{self-similar} wake in the outer disc. More specifically, we now assume the surface density perturbation due to the density wave to have the form
\begin{align}
\delta \Sigma(R,\phi) = \delta \tilde\Sigma(R)  \times\psi\left(\phi - \tilde\phi(R) \right),
\label{eq:ansatz}
\end{align}
where function $\psi$ describes the azimuthal structure of the wake, while the pre-factor $\delta \tilde\Sigma(R)$ describes the evolution of the wave amplitude (equation (\ref{eq:Gauss}) has such a form). In this self-similar picture, the wake travels with varying amplitude along the curves $(R,\tilde\phi(R))$, and the only change of $\psi$ with $R$ is a translation in the $\phi$-direction given by $\tilde\phi(R)$. 

With the ansatz (\ref{eq:ansatz}) we have
\begin{align}
\pd{\delta\Sigma}{\phi} &= \delta\tilde\Sigma(R)\times
\psi^\prime(\phi-\tilde\phi(R)),
\label{eq:deriv1}
\end{align}
where $\psi^\prime(x)=\de\psi(x)/\de x$.

Let us now introduce the complex Fourier coefficients of $\psi'$ as follows:
\begin{align}
    \Psi_m &\equiv \frac{1}{\pi}
             \int_0^{2\pi}
             \psi' (x) \exp (-\I m x )
             \, \de x,~~~~~m=1,...,\infty.
    \label{eq:def_psim}
\end{align}
Writing $\Psi_m = A_m e^{\I \theta_m}$, with $A_m = \vert \Psi_m \vert$ and $\theta_m = \arg \Psi_m$ being the amplitude and the phase of $\Psi_m$, respectively, we obtain
\begin{align}
\frac{\partial\delta\Sigma}{\partial \phi}  &=
\delta\tilde\Sigma(R)
\sum_{m=1}^\infty
    \RE\left\lbrace
    \Psi_m \exp[\I m (\phi-\tilde\phi(R))]
    \right\rbrace,\\
    \label{eq:fs_psi}
&=
\delta\tilde\Sigma(R)
\sum_{m=1}^\infty
    A_m \cos \left[ m (\phi-\tilde\phi(R)) + \theta_m \right],
\end{align}
Substituting this expression into equation (\ref{eq:dT2}) gives, after straightforward manipulation, 
\begin{align}
    \Tex = - \pi G \Mp \, \delta\tilde\Sigma(R) \sum\limits_{m=1}^{\infty} A_m \, b_{1/2}^{(m)}(\alpha) \cos[m \tilde\phi(R) - \theta_m].
    \label{eq:dT4}
\end{align}
Note that here $\tilde\phi(R)$ is understood to lie in the interval $(0,2\pi)$, i.e. is unique mod $2\pi$, since in equation (\ref{eq:deriv1}) $\phi\in (0,2\pi)$.

This is the final expression for the torque density given the self-similar ansatz (\ref{eq:ansatz}). Once the forms of $\delta\Sigma(R)$, $\tilde\phi(R)$ and $\Psi_m$ (i.e. $A_m$ and $\theta_m$) are specified, the torque density can be computed by simple summation. Knowledge of $\Psi_m$ is crucial for this calculation and in Section \ref{sec:modal} we discuss the properties of these coefficients for a certain class of disc models described next.


\subsection{Self-similarity in AMF-preserving discs}
\label{sec:planet_waves}


\citet{Miranda2019I} have shown that the self-similar ansatz (\ref{eq:ansatz}) works well in the outer parts of the AMF-preserving (adiabatic) discs: the outer density wave maintains a single-armed, close to self-similar shape, see Fig. 4a of that paper that reveals only a weak evolution of the wake shape for $R>\Rp$. This is also obvious from our Fig. \ref{fig:tq_map}c drawn for the fiducial disc model: azimuthal profiles of the wake at different radii look essentially identical (once one shifts each profile horizontally). 

At the same time, the self-similar ansatz is clearly not applicable in the inner disc as a result of secondary (and tertiary) arm development there, see Figs. 4b,5,6 of \citet{Miranda2019I}. Our Fig. \ref{fig:tq_map}d also makes this clear since the wake profiles for different $R$ have noticeably different shape and no horizontal shift would make them match. 

This picture remains largely unchanged as one varies disc parameters, as we demonstrate in Section \ref{sec:wake_pars}. Given all that, in the rest of this section we will focus on planet-driven density waves in the outer regions of AMF-preserving discs (with not too low $\hp$, see Section \ref{sec:res_var_hp}), for which the ansatz (\ref{eq:ansatz}) works well.

By definition, the angular momentum of the freely-propagating (i.e. not subject to further forcing) density waves is conserved in the AMF-preserving (e.g. adiabatic) disc models, i.e. $\partial F_J/\partial R=0$. It has been shown \citep{R02} that conservation of $F_J$ leads to the wave amplitude scaling as $\delta \Sigma(R) \propto \dslin(R)$, where
\begin{align}
    \dslin &= \Sigma(R) \frac{\Mp}{\Mth}
    \left(  \frac{\vert \Omega(R) - \Omegap \vert}{\Omegap} \frac{\Sigma_\mathrm{p}}{\Sigma(R)} \frac{\Rp}{R} \right)^{1/2}
    \left( \frac{c_\mathrm{p}}{\cs (R)} \right)^{3/2}.
    \label{eq:dslin}
\end{align}
In a power-law background disc, see equations (\ref{eq:T_i}) \& (\ref{eq:Sig_i}), we find\footnote{The $\vert (R/\Rp)^{-3/2} - 1 \vert$ factor is missing the absolute magnitude in equations (37) of \citet{R02} and  (16) of \citet{Miranda2019I}. It is correctly present in the function $g(R)$ in \citet{R02}.} 
\begin{align}
    \dslin \propto \left\vert (R/\Rp)^{-3/2}-1 \right\vert^{1/2} \left(\frac{R}{\Rp}\right)^{(3q/4 - p/2)}.
    \label{eq:dslin2}
\end{align}

It is important to emphasize that the result (\ref{eq:dslin}) would not be valid in the non-AMF-preserving discs. For example, in discs with locally isothermal EoS $F_J$ varies as $F_J \propto \cs^2(R)$ \citep{Lin2015,Miranda2019I}, so that the wave amplitude gets amplified (damped) relative to $\dslin$ in the inner (outer) disc. This can be easily accounted for by changing the power of $c_\mathrm{p}/\cs(R)$ in equation (\ref{eq:dslin}) from $3/2$ to $1/2$. Similarly, thermal relaxation (e.g. $\beta$-cooling) always damps $F_J$ \citep{Miranda2020I,Miranda2020II}, reducing $\delta \Sigma(R)$ relative to $\dslin$; in this case no simple analytical correction to equation (\ref{eq:dslin}) is possible. We will discuss the impact of these alternative (to adiabatic) thermodynamic assumptions in Section \ref{sec:res_beta}.

The discussion above implies that the density waves in the outer parts of the AMF-preserving discs should be well approximated by the ansatz (\ref{eq:ansatz}) with
\begin{align}
\delta \tilde\Sigma(R) = \delta \Sigma_{\rm lin}(R),~~~~  \tilde\phi(R)=\phil(R),
\label{eq:ansatz1}
\end{align}
with $\phil(R)$ given by equations (\ref{eq:phi-lin})-(\ref{eq:lx}).


\subsection{Properties of $\Psi_m$ in AMF-preserving discs}
\label{sec:modal}


\begin{figure}
    \begin{center}
    \includegraphics[width=0.49\textwidth]{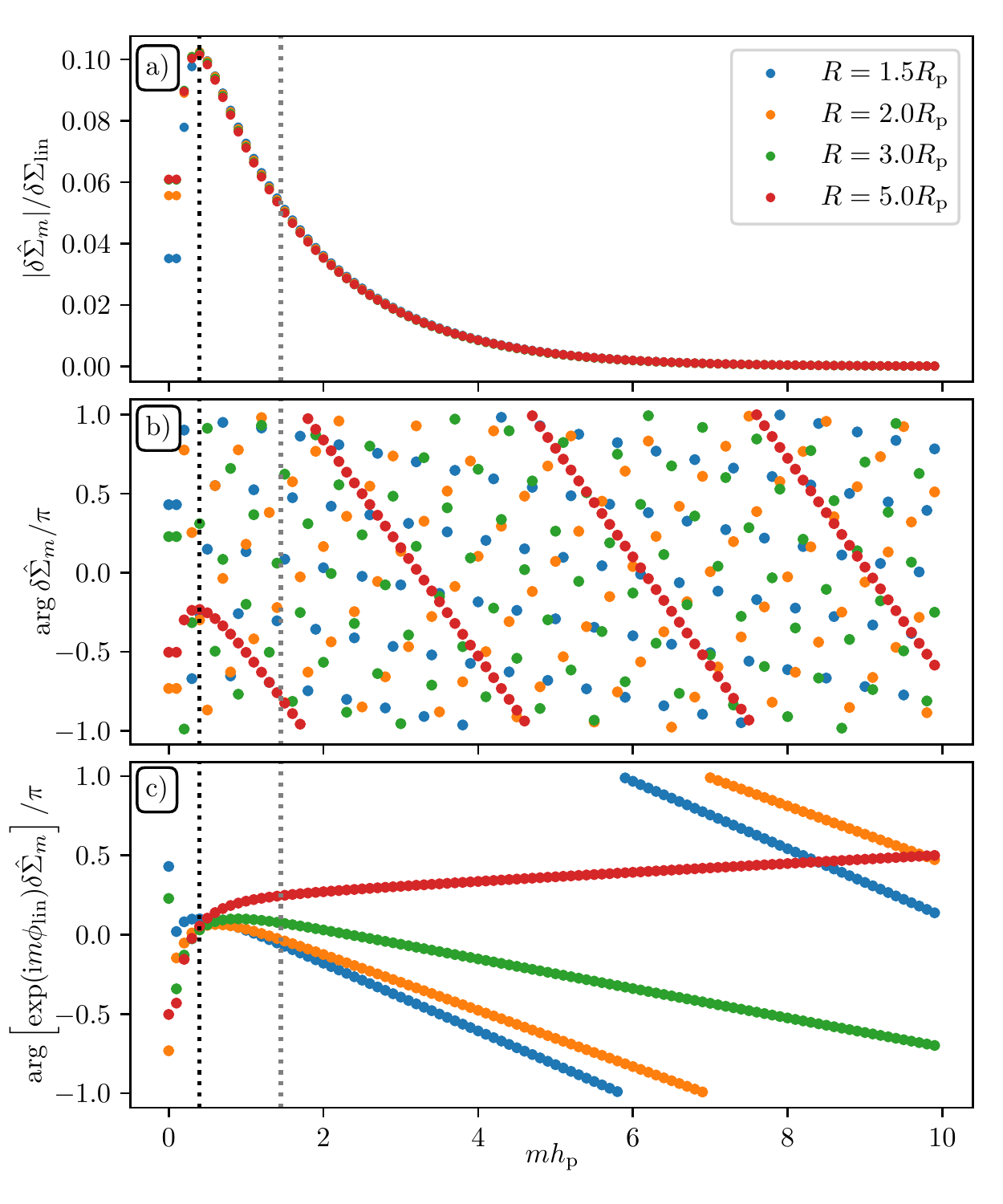}
    \vspace*{-.6cm}
    \caption{
        Amplitudes (a) and phases (b) of  $\delta \hat{\Sigma}_m$ ---the Fourier components of the surface density perturbation in the outer disc; panel (c) shows phases with an additional shift by $\phil$(R), see equations (\ref{eq:phi-lin}), (\ref{eq:Deltaphi}). The grey dotted vertical line corresponds to the same radius as in Fig. \ref{fig:amp_phase_fid}, where $A_m$ peak and the black dotted line show where $\vert \delta \hat{\Sigma}_m \vert$ peak ($m \hp \simeq 0.4$). This plot illustrates the radial universality of amplitudes and evolution of phases of the planet-driven density wave. See Section \ref{sec:modal} for details.
    }
    \label{fig:four_dsig_amp_phase}
    \end{center}
\end{figure}


\begin{figure*}
    \begin{center}
    \includegraphics[width=\textwidth]{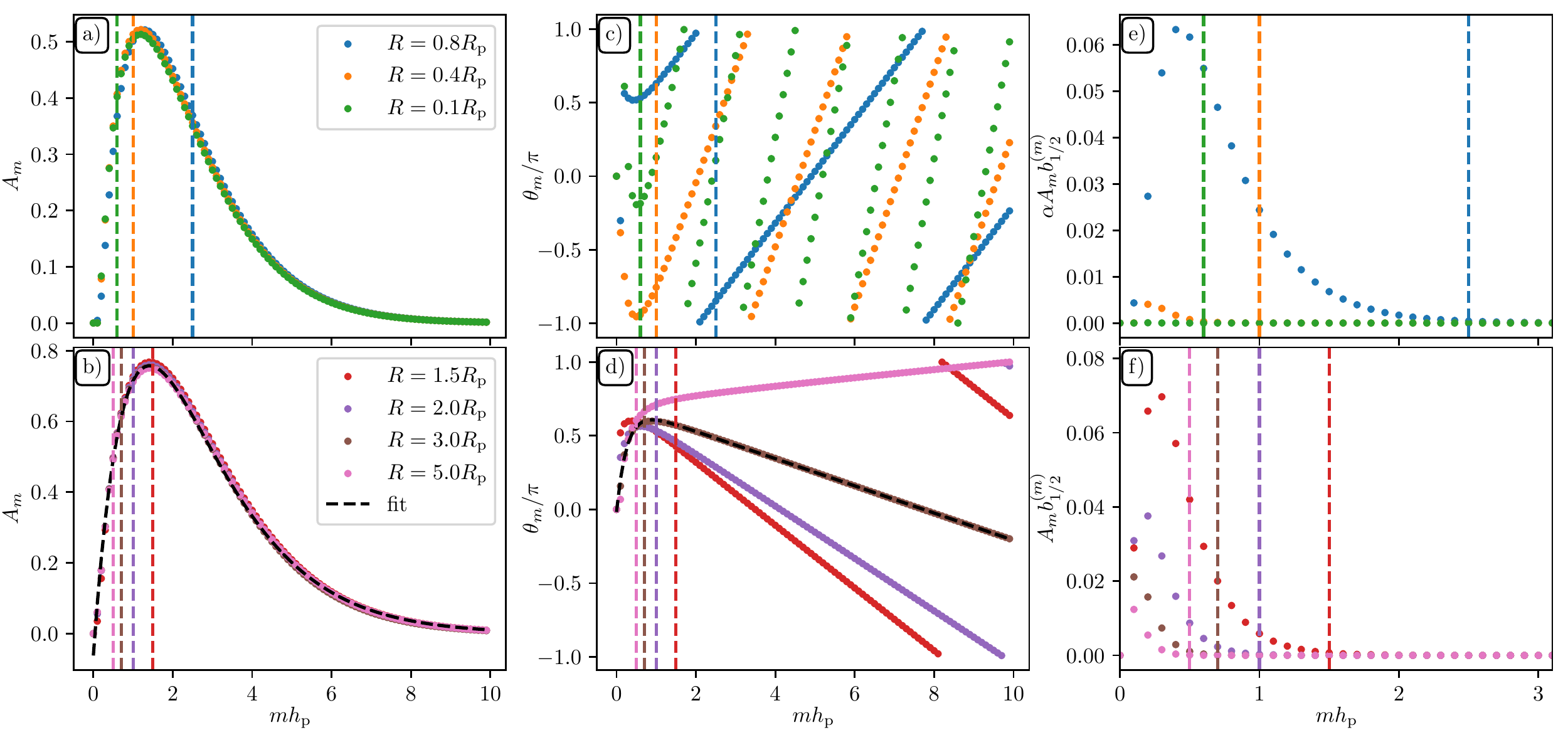}
    \vspace*{-.6cm}
    \caption{
        Amplitude of $\Psi_m$ ($A_m$, left), phase of $\Psi_m$ ($\theta_m$, middle) and $A_m$ multiplied by the corresponding Laplace coefficient (right) and $\alpha$ in panel (e) as a function of $m \hp$ for the fiducial disc for several radii (colours) in the inner (top) and outer (bottom) disc. The dashed vertical lines indicate $m_\mathrm{c} \hp$, such that only the modes with $m <m_\mathrm{c}$ (to the left of those lines) contribute significantly to $\Texs$. See text for details. 
    }
    \label{fig:amp_phase_fid}
    \end{center}
\end{figure*}

Next, we explore the properties of the Fourier coefficients $\Psi_m$ for the self-similar outer density waves in AMF-preserving discs. We employ the ansatz (\ref{eq:ansatz}), (\ref{eq:ansatz1}) and consider an adiabatic disc with $\hp =0.1$ and $p=q=1$.

The Fourier coefficients of $\psi$ can be obtained via the Fourier coefficients of $\delta \Sigma$, which are available to us from the solution of the linear problem. They are given by
\begin{align}
    \delta \hat{\Sigma}_m (R) &\equiv \frac{1}{\pi}
                                \int_0^{2\pi}
                                \Sigma(R,\phi) e^{-\I m \phi}
                                \, \de \phi,~~~~~~\mbox{such that}
                                \label{eq:coeff_dsig}\\
    \delta \Sigma(R,\phi) &= \sum_\mathrm{m=1}^{\infty}
                            \RE
                            \left\lbrace \delta \hat{\Sigma}_m (R) e^{\I m \phi } \right\rbrace.
    \label{eq:fs_dsig}
\end{align}
Using equation (\ref{eq:fs_psi}) one can easily show that
\begin{align}
    \Psi_m (R) = \I m\frac{\delta \hat{\Sigma}_m (R)}{\dslin(R)} e^{\I m \phil(R)}.
    \label{eq:rel_psi_dsig}
\end{align}
In particular, the absolute magnitude of the Fourier coefficients of $\psi$ and $\delta \Sigma$ are related by $\vert \Psi_m \vert = m  \vert \delta \Sigma_m \vert/\dslin$. 

In Fig. \ref{fig:four_dsig_amp_phase} we show the behaviour of the amplitude and phase of $\delta \hat{\Sigma}_m$ for several values of $R$, which is helpful for understanding the properties $\Psi_m$ later on. One can see in panel (a) the almost universal shape of $\vert\delta \hat{\Sigma}_m (R)\vert$, weakly dependent on $R$ and consistent with the self-similar description of $\delta\Sigma$. On the other hand, panel (b) looks like a scatter plot and shows that the phases of $\delta \hat{\Sigma}_m (R)$ (computed in a frame with azimuthal axis aligned with $\phip$) do not follow a clear pattern while varying rapidly. However, once we shift these phases by $\phil(R)$ to account for the overall wrapping of the spiral wake (panel (c)), a clear pattern becomes obvious, showing only a slow evolution with $R$. This again supports the overall picture of a self-similar wake and shows that it closely follows the predicted path.

Next we turn to examining the amplitude $A_m=\vert\Psi_m\vert$ and phase $\theta_m=$ arg$\,\Psi_m$ of the coefficients $\Psi_m$ computed using $\delta \hat{\Sigma}_m$ via equation (\ref{eq:rel_psi_dsig}). In Fig. \ref{fig:amp_phase_fid} we show them as a function of $m \hp$, in the left and middle column, respectively, both in the outer disc (bottom) and, for completeness, in the inner disc (top).

Looking at the left column (panels a \& b), we notice that $A_m$ peak at around $m \hp \simeq 1.45$ ($m \hp \simeq 1.2\Rp$) in the outer (inner) disc and do not show much variation with $R$. This is expected for free linear waves in {\it flux-conserving} (adiabatic and barotropic) disc models, since the radial scaling of $\delta\Sigma$ with $\dslin(R)$ is already taken into account in ansatz (\ref{eq:ansatz}), (\ref{eq:ansatz1}). As we show in Section \ref{sec:res_beta}, the situation is very different for non-flux-conserving discs.

On the other hand, the phases $\theta_m$ (panels c \& d), which include the phase shift by $\phil(R)$ through equation (\ref{eq:rel_psi_dsig}), as in Fig. \ref{fig:four_dsig_amp_phase}c, show a substantial evolution with $R$. In the inner disc, the phases spread out with $m$ as the distance from the planet increases. Around $R = 0.1 \Rp$,  $\theta_m$ changes by $2 \pi$ every $\Delta (m \hp) \simeq 2$. On the other hand, in the outer disc, $\theta_m$ change a lot less with $m$ and, in particular, show very weak variation with $R$ at low $m$ (similar to Fig. \ref{fig:four_dsig_amp_phase}c). 

Panel (f) shows the amplitudes $A_m$ multiplied with Laplace coefficients $\lpm(\alpha)$, to illustrate the contribution of different azimuthal harmonics of $\delta\Sigma$ to the torque density in the outer disc, see equation (\ref{eq:dT4}). Note that we show these over a smaller range of $m \hp$, in order to zoom into the relevant region. For $R$ closest to the planet, $A_m b_{1/2}^{(m)}$ peak at around $m \hp \simeq 0.4-0.5$ and fall off rapidly at large $m$. As the distance from the planet increases, the location of the peak shifts towards lower $m$, with the overall amplitude of $A_m b_{1/2}^{(m)}$ decaying. This is expected from the asymptotic behaviour of the Laplace coefficients (discussed in Section \ref{sec:asy_analysis}) and the weak dependence of $A_m$ on $R$. In other words, the behaviour of the Laplace coefficients is the key factor determining which $m$ provides the largest contribution to $\Texs$. Vertical dashed lines in all panels of Fig. \ref{fig:amp_phase_fid} correspond to $m_\mathrm{c}$ defined as the critical $m$ for which the product $A_m \lpm$ falls off below one percent of its maximum value, such that only $m < m_\mathrm{c}$ need to be considered when computing $\Texs$ (the values are $m_\mathrm{c} = 14, 9, 6, 4$ for $R = 1.5, 2, 3, 5\Rp$, respectively.). These vertical dashed lines move to lower $m$ as $\vert R-\Rp\vert$ increases, implying that closer to the planet, more modes contribute to $\Texs$, while for larger distances, only the lowest few $m$ matter; see Section \ref{sec:asy_analysis} for more details.

Similarly, in the inner disc we multiply $A_m$ with $\alpha \lpm(\alpha)$ (panel (e)), see equation (\ref{eq:dT3_in}). Because of the extra factor of $\alpha$ (compared to the outer disc) the resultant curves precipitously diminish in magnitude as $R$ decreases, explaining the weakness of wiggles in the inner disc. Dashed lines again mark $m_c$, for which $A_m\alpha \lpm(\alpha)$ fall to $1\%$ of the maximum value.

The behaviour of the phases $\theta_m$ in panel (c) makes it obvious that self-similarity is not a good approximation in the inner disc, consistent with the left columns of Figs. \ref{fig:psi_var_pq} \& \ref{fig:psi_var_h}. But the variation of $\theta_m$ at high $m$ in panel (d) suggests that self-similarity may not work in the outer disc either. However, note that in panel (d) the variation of $\theta_m$ with $R$ is not very dramatic for $m<m_\mathrm{c}$ (to the left of the corresponding vertical lines), i.e. for all azimuthal harmonics that matter for the calculation of $\Texs$. This allows us to consider $\delta\Sigma$ as {\it approximately} self-similar in the outer disc, which will be used next. 


\begin{figure}
    \begin{center}
    \includegraphics[width=0.49\textwidth]{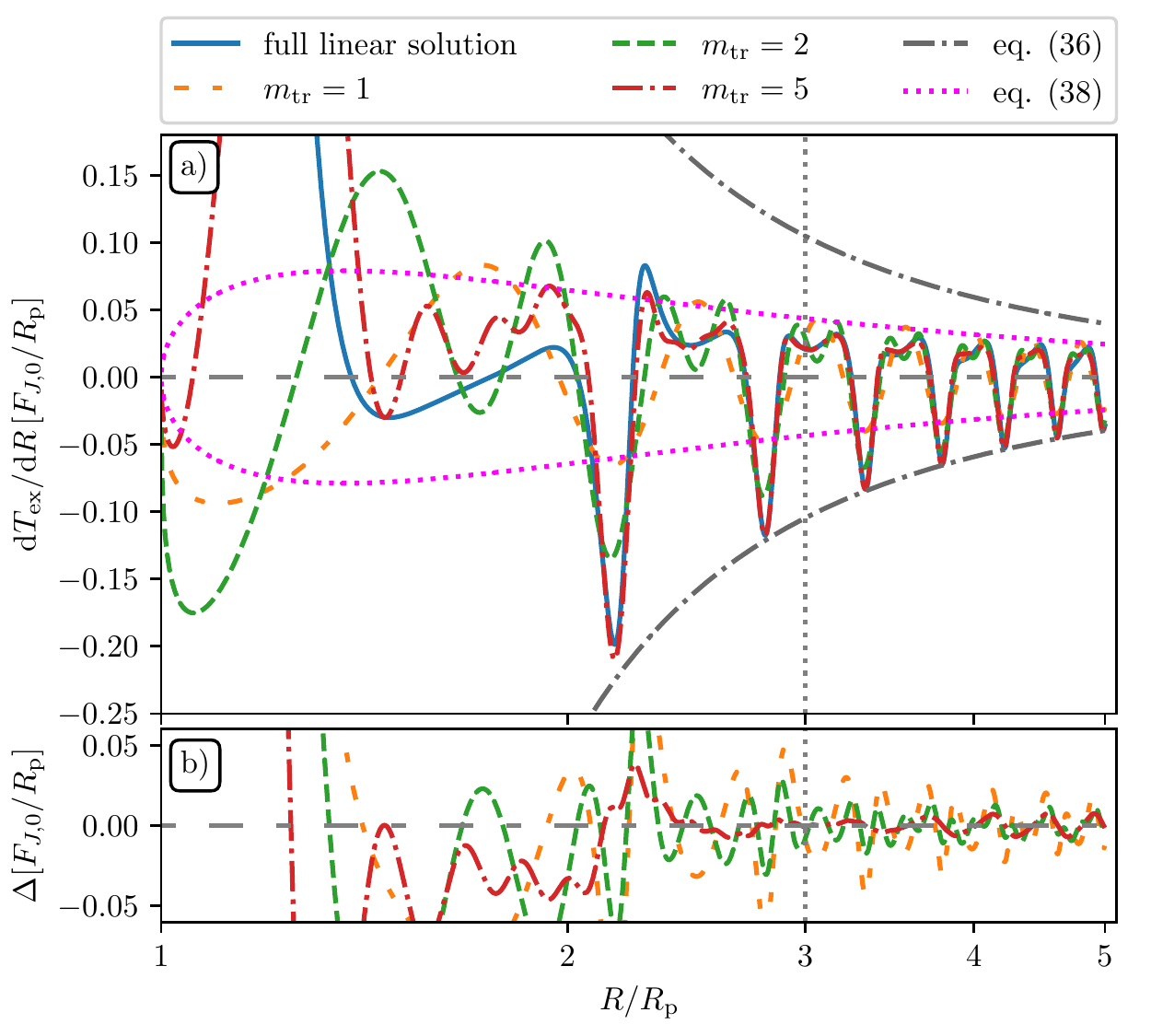}
    \vspace*{-2em}
    \caption{
        Illustration of the convergence of the series  (\ref{eq:truncseries_dtdr}) to the exact expression (\ref{eq:dT4}).
        {\it Top:}
        The blue solid line shows the full linear solution using $\mmax=100$ modes with $(A_m,\theta_m)$
        varying with radius (see Fig. \ref{fig:amp_phase_fid}), i.e. without using the self-similar approximation. Dashed lines show the approximation (\ref{eq:truncseries_dtdr}) truncated at different $\mtr < \mmax$ (colours) with
        $(A_m,\theta_m)$ sampled at $\Rcal=3\Rp$  (see Fig. \ref{fig:amp_phase_fid}). Dot-dashed grey and dotted magenta curves illustrate two predictions for the envelope of the torque wiggles given by the equations (\ref{eq:dTas1}) and (\ref{eq:dTas}), respectively.
        {\it Bottom:} deviation $\Delta$ of the truncated series (\ref{eq:truncseries_dtdr}) from the exact linear solution. Agreement is best around $\Rcal$ and the truncated series rapidly converge with $\mtr$ to the exact solution, especially at larger $R$. 
    }
    \label{fig:conv_series_fid}
    \end{center}
\end{figure}



\subsection{Semi-analytic reconstruction of torque wiggles in AMF-preserving discs}
\label{sec:rec_tex_fid}


Now we demonstrate how the approximate self-similarity of the density perturbation in the outer disc allows us to reconstruct the main features of the torque wiggles behaviour. We compute $\Texs$ using equation (\ref{eq:dT4}) with $A_m$, $\theta_m$ obtained under the assumption of a self-similar wake, see Section \ref{sec:modal}. Since Fig. \ref{fig:amp_phase_fid}d shows some evolution of $\theta_m$ with $R$, we pick a particular {\it calibration} radius $\Rcal=3$, at which we measure $A_m$, $\theta_m$ and assume these values to not depend on $R$ (an approximation which should work reasonably well as we argued earlier) when computing the radial profile of $\Texs$. The resultant torque density is then compared with the $\Texs$ obtained by solving the linear problem exactly. 

In order to highlight the contribution of different individual modes to $\Texs$, we also use a truncated (at $m=\mtr$) version of the series expansion (\ref{eq:dT4}):
\begin{align}
    \nonumber
    \Textr(\mtr,\Rcal,R)= &- \pi G \Mp \dslin(R) \sum\limits_{m=1}^{\mtr}
    b_{1/2}^{(m)} (\alpha) \\
    &\times A_m(\Rcal) \cos[m \phil (R) - \theta_m(\Rcal)],
    \label{eq:truncseries_dtdr}
\end{align}
where here we have made explicit that under our self-similar approximation the values of $(A_m,\theta_m)$ are taken at a calibration radius $\Rcal$.

In Fig. \ref{fig:conv_series_fid} we show the exact numerical solution for $\Texs$ (blue solid) and the truncated self-similar series $\Textrs$ (\ref{eq:truncseries_dtdr}) for several values of $\mtr$ in panel (a); the deviation between them is shown in panel (b). One can see that close to the planet ($R \leq 2\Rp$) the self-similar approximation deviates significantly from the global linear solution for all $\mtr$. This is to be expected since $\theta_m$ still show some dependence on $R$ (compared to their values at $\Rcal$) for $m\lesssim m_\mathrm{c}$ close to the planet, see Fig. \ref{fig:amp_phase_fid}d. Further away, the first 
trough of $\Texs$ is matched well both in shape and amplitude for all $\mtr > 2$. In the radial interval between this trough and the next one, we see rather small deviations for all $\mtr$. Beyond the second trough, the self-similar curves closely track the full solution for
$\mtr \geq 2$. For $R \geq 4 \Rp$, the curve for $\mtr = 2$ merges with those for higher $\mtr$. The outermost torque oscillation period is matched reasonably well even for a single mode when $\mtr = 1$. Panel (b) reveals that for $R \geq 4\Rp$, the self-similar approximation shows a periodic deviation from the exact solution, with the amplitude that steadily decays as $\mtr$ increases.

These results indicate that for sufficiently large distances from the planet, $\mtr = 5-10$ modes are sufficient to describe the behaviour of $\Texs$, if the $\Psi_m$ are sampled at an intermediate distance $\Rcal$ within the radial region of interest. At the quantitative level these results still depend on the value of $\Rcal$, but this dependence should not curtail the good performance of the self-similar approximation described in Sections \ref{sec:selfim},\ref{sec:planet_waves} in reproducing $\Texs$.


\subsection{Understanding the key features of torque wiggles}
\label{sec:asy_analysis}


As demonstrated in the previous subsection, equation (\ref{eq:dT4}) is able to successfully reproduce the correct behaviour of $\Texs$. Armed with this knowledge, we now use it to understand some key properties of the torque wiggles. 

First, equation (\ref{eq:dT4}) allows us to predict reasonably well the shape of the {\it outer envelope} of the oscillating torque wiggles. This can be done by setting the cosine factors in equation (\ref{eq:dT4}) to unity, thus allowing the expression in the right hand side its maximum possible value
\begin{align}
    {\rm max} \left|\Tex\right|  = \pi G \Mp \, \delta\tilde\Sigma(R) \sum\limits_{m=1}^{\infty} b_{1/2}^{(m)}(\alpha) A_m.
    \label{eq:dTas1}
\end{align}
This very simple constraint is plotted as the grey dot-dashed curve in Fig. \ref{fig:conv_series_fid}a and it works remarkably well, with the troughs of peak wiggles almost touching this curve (see more on this below).

Second, equation (\ref{eq:dT4}) lets us analyze the torque density behaviour in the asymptotic limit $R\gg\Rp$ (or $\alpha\ll 1$). Recall that in thin discs, $H/R\ll 1$, the Fourier amplitudes of planet-driven $\delta\Sigma_m$ peak at $m_* \sim \Rp/\Hp$ (i.e. $m_* \hp \sim 1$) and exponentially decay for $m\gtrsim m_*$ \citepalias{GT80}. The coefficients $\Psi_m$ have an extra factor of $m$ compared to $\delta\Sigma_m$, leading to $A_m$ peaking at $m$ slightly higher than $m_*$, which can be seen by comparing Figs. \ref{fig:four_dsig_amp_phase} (see black and grey dotted lines) and \ref{fig:amp_phase_fid}. When computing $\Texs$ via formula (\ref{eq:dT4}), $A_m$ are multiplied by the Laplace coefficients $\lpm$, which asymptotically behave far from the planet as \citep{MurrayDermott}
\begin{align}
    b_{1/2}^{(m)} (\alpha) \propto \alpha^m,~~~\alpha=\Rp/R \ll 1.
    \label{eq:asy_lapl}
\end{align}
This suggests that as $\vert R - \Rp \vert$ increases, the main contributions to $\Texs$ should be provided by the low-$m$ components of $\delta\Sigma$. Eventually, far from the planet, only the $m=1$ component would matter. In other words, we expect a transition from the interference of many modes closer to the planet to a single-mode dominated behaviour far away, just as we observed in Section \ref{sec:rec_tex_fid}.

Retaining only the first term in the sum in equation (\ref{eq:dT4}) and using the fact that $b_{1/2}^{(1)} (\alpha) \to \alpha$ as $\alpha\ll 1$, we obtain the following asymptotic prediction for the outer envelope of $\Texs$:
\begin{align}
    {\rm max} \left|\Tex\right|  \approx \pi G \Mp \, A_1\frac{\Rp}{R}\,\delta\Sigma_\mathrm{lin}(R),
    ~~~~~~~R\gg\Rp,
    \label{eq:dTas}
\end{align}
where the dependence of $\delta\Sigma_\mathrm{lin}$ on $R$ is given by the equation (\ref{eq:dslin}). This prediction is illustrated in Fig. \ref{fig:conv_series_fid} via the dotted magenta curves and it should be working well at large $R\gg \Rp$; at intermediate $R$ it underestimates the envelope (\ref{eq:dTas1}) of $\Texs$ because of still significant contribution of other $m$ harmonics.

Note that in the inner disc close to the centre, in the asymptotic limit $\alpha=R/\Rp\to 0$, torque density is suppressed by an additional factor of $\alpha$: $\Texs \sim \delta\Sigma_1 (R/\Rp)^2$, where $\delta\Sigma_1$ is the $m=1$ Fourier component of $\delta\Sigma$, see the equation (\ref{eq:dT3_in}). As a result, torque contributions of different harmonics fall off very rapidly as $R$ decreases, see Fig. \ref{fig:amp_phase_fid}e. This, together with the decoherence of the planetary wake due to the formation of multiple spirals, acts to suppress the torque wiggles for $R<\Rp$. 

Third, equation (\ref{eq:dT4}) also allows us to explain why in the outer disc $\Texs$ features sharp (negative) peaks at radii, where the density wake crosses the line $\phi=0$ between the star and the planet. Indeed, note that $\theta_m$ in Fig. \ref{fig:amp_phase_fid}d exhibit rather small spread $\Delta\theta_m\lesssim \pi/2$ for all $R$ and $m\le m_\mathrm{c}$. This allows us to assume that $\theta_m$ is approximately the same for all $m$ that contribute significantly to the sum in equation (\ref{eq:dT4}). In that case, these significant terms {\it interfere constructively} whenever $m\phi_\mathrm{lin}(R)\lesssim 1$ for all $m\lesssim m_\mathrm{c}$ (this also explains why the troughs of $\Texs$ overshoot the envelope (\ref{eq:dTas}) at intermediate $R$). As a result, whenever $\phi_\mathrm{lin}(R)\to 0$ mod $2\pi$, i.e. the wake (centred on $\phi_\mathrm{lin}(R)$) crosses the line $\phi=0$ from the star to the planet, constructive interference of many terms in (\ref{eq:dT4}) leads to a sharp increase of the amplitude of $\Texs$, producing the characteristic appearance of the torque wiggles. In fact, the interference at these locations is so effective, that the trough wiggles (negative values) almost reach the maximum possible absolute value of $\vert\Texs\vert$ given by equation (\ref{eq:dTas1}).

Fourth, we can also use equation (\ref{eq:dT4}) to estimate the radial width of these sharp features of $\Texs$ (although the outcome of this exercise may be of limited utility). Indeed, constructive interference of harmonics with $m\lesssim m_\mathrm{c}$ leading to torque wiggles requires $\delta \phi_\mathrm{lin}\lesssim m_\mathrm{c}^{-1}$, constraining the radial width of the sharp troughs of $\Texs$ as
\begin{align}
\delta R
    \sim \delta \phi_\mathrm{lin} \left(\frac{\partial\phi_{\rm lin}}{\partial R}\right)^{-1}
    \lesssim m_\mathrm{c}^{-1} \frac{\cs(R)}{\Omegap-\Omega(R)},
 \label{eq:dr}
 \end{align}
see equation (\ref{eq:Deltaphi}). Because of the rapid evolution of $m_\mathrm{c}$ with $R$ (see Fig. \ref{fig:amp_phase_fid}) it may be somewhat challenging to extract a more explicit dependence of $\delta R$ on $R$ and disc parameters. Nevertheless, it is clear that $\delta R$ should be smaller in thinner discs with lower $\cs$ and $\hp$.

\begin{figure}
    \begin{center}
    \includegraphics[width=0.49\textwidth]{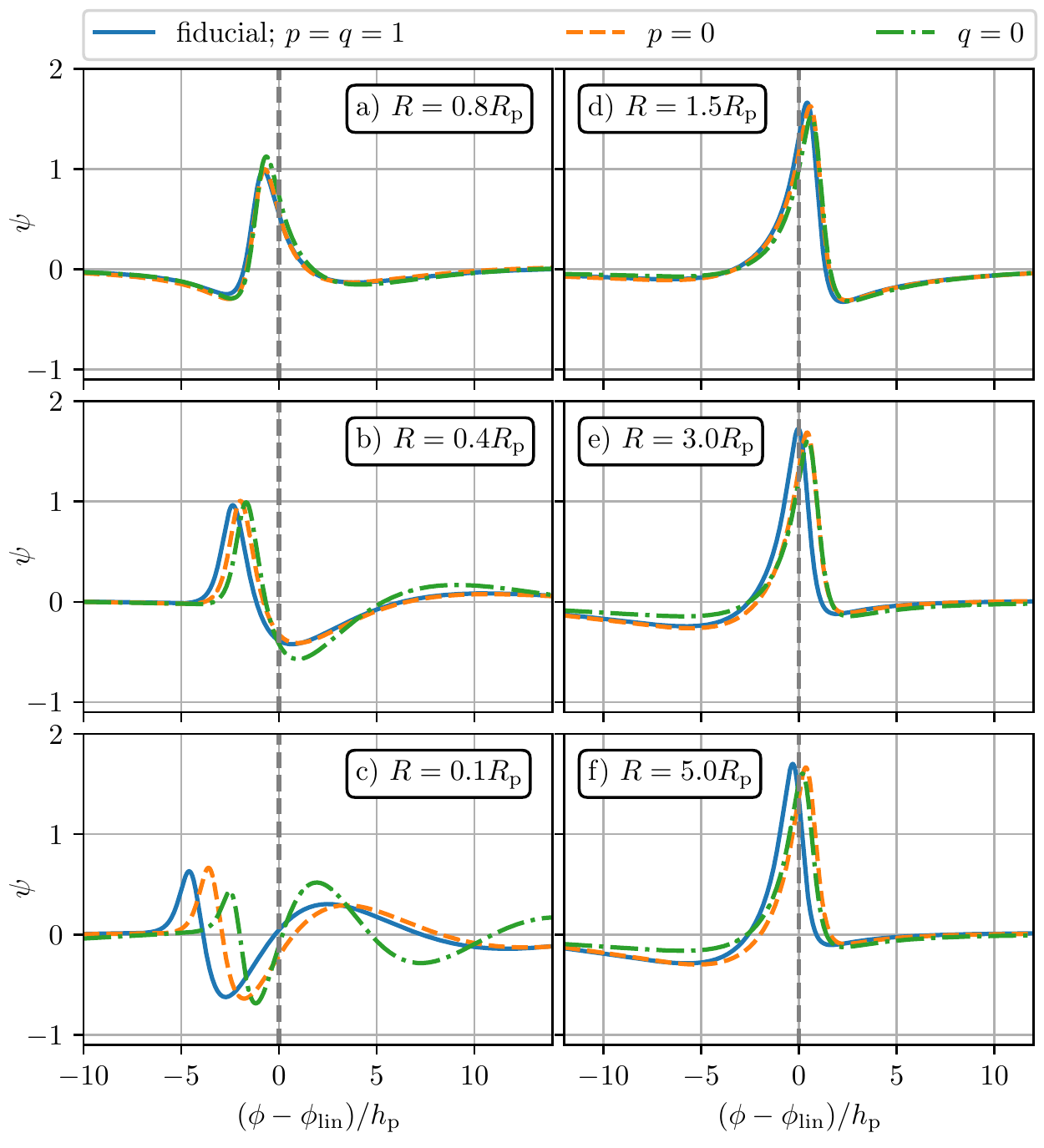}
    \vspace*{-0.5cm}
    \caption{
    The dimensionless function $\psi(x)$ describing the azimutal shape of the wake in the self-similar anzatz (\ref{eq:ansatz}) sampled at different $R$ (panels) in the inner disc (left column) and outer disc (right column) for different values of the surface density and temperature slopes $p$ and $q$ (colours).
    The coordinate on the $x$-axis has been rescaled $\propto \hp^{-1}$. See text for discussion.
    }
    \label{fig:psi_var_pq}
    \end{center}
    \end{figure}
\begin{figure}
    \begin{center}
    \includegraphics[width=0.49\textwidth]{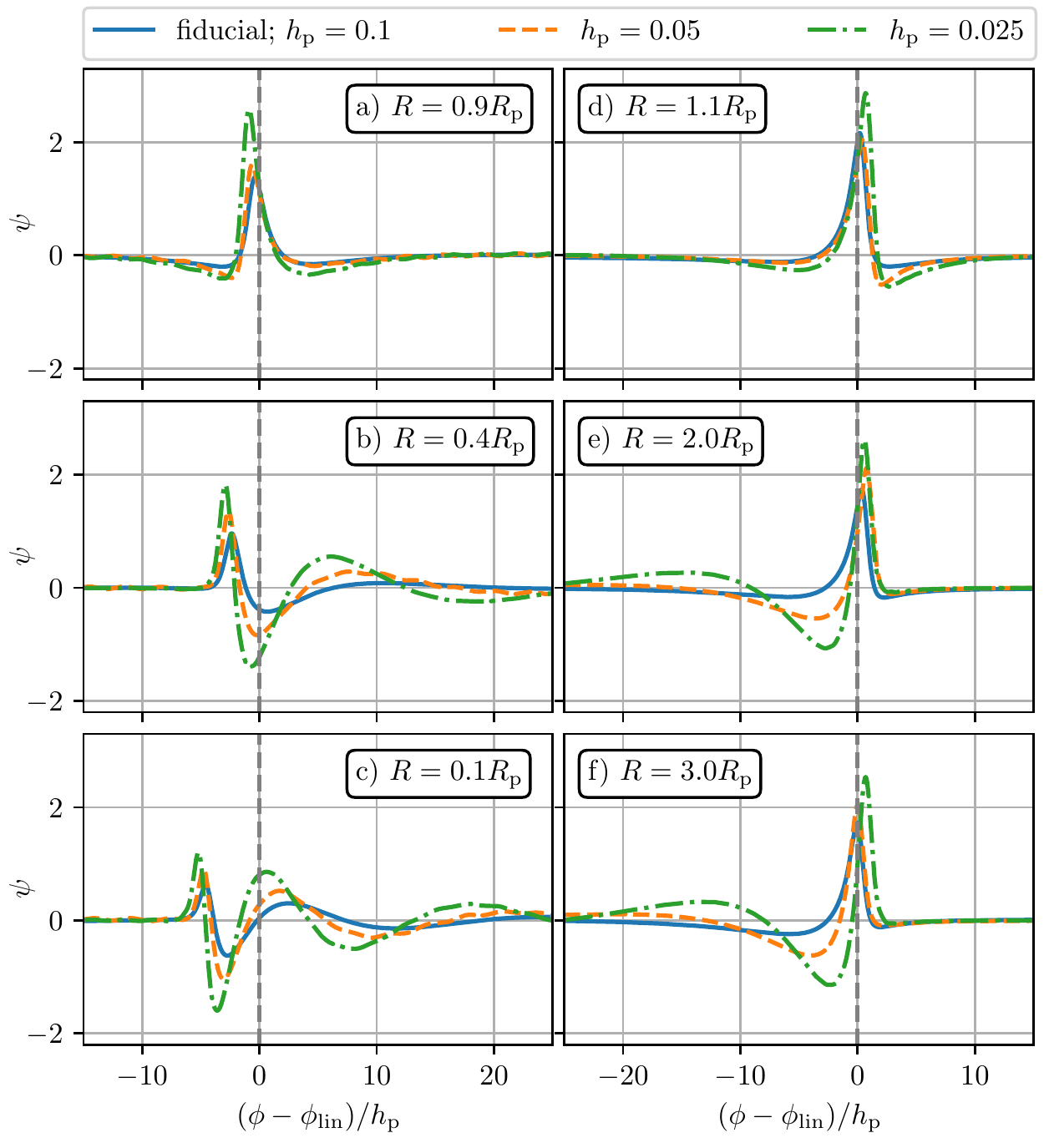}
    \vspace*{-0.5cm}
    \caption{Same as Fig. \ref{fig:psi_var_pq} but now illustrating the effect of the disc scaleheight $\hp$ variation on the wake profile evolution. The coordinate on the $x$-axis has been rescaled $\propto \hp^{-1}$, such that the wakes have similar horizontal scale. Note that for lower $\hp$ the dispersion and splitting of the wave accelerate and self-similarity gets violated, even in the outer disc. See text for details.
    }
    \label{fig:psi_var_h}
    \end{center}
\end{figure}

\begin{figure*}
    \begin{center}
    \includegraphics[width=0.99\textwidth]{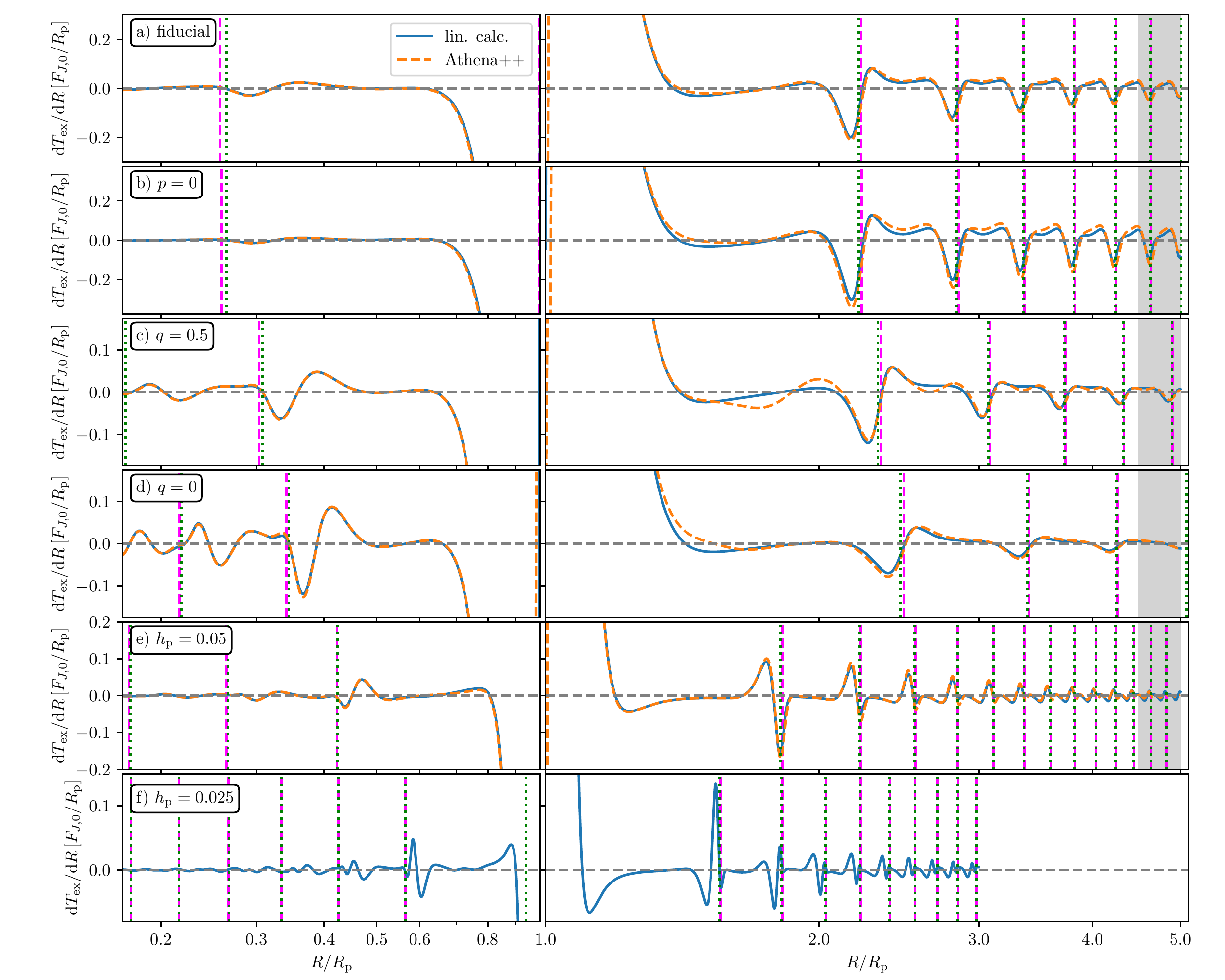}
    \vspace*{-.1cm}
    \caption{Torque density profiles obtained using linear theory (blue), and direct  Athena++ simulations with $\Mp = 0.01 \Mth$ (orange), for different disc parameters (the value of the parameter varied relative to the fiducial model is indicated in each panel). Vertical magenta (dashed) lines correspond to the radial locations where the peak of the wake crosses the line $\phi=\phip$ from the star to the planet, while the green (dotted) lines correspond to the analytical prediction $\phil=2\pi n$, $n=1,2,...$. For the smallest value $\hp = 0.025$, no Athena++ simulation was performed. Note that axes are broken and log-scale is different on left and right hand side. Grey shaded areas indicate damping zones in the outer disc in Athena++ simulations (inner disc damping zones are outside of plot range).
    }
    \label{fig:dTdR_summary}
    \end{center}
\end{figure*}
 
\begin{figure}
\begin{center}
\includegraphics[width=0.49\textwidth]{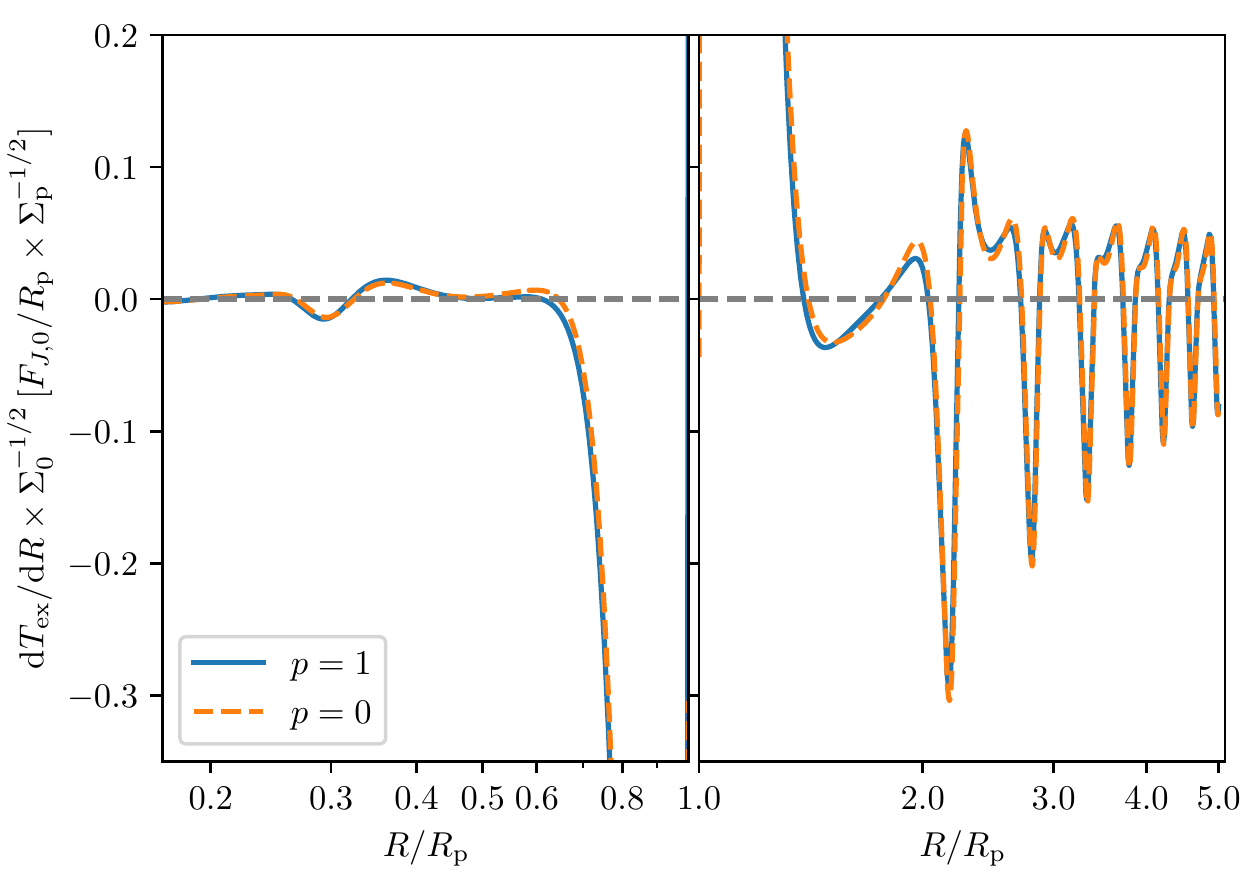}
\vspace*{-0.4cm}
\caption{Comparison of torque densities, divided by $\Sigma_0^{1/2}$, to eliminate the scaling of $\Texs$ with $\delta \Sigma_\mathrm{lin}$ for fiducial $p=1$ (solid) and for $p=0$ (dashed). This illustrates that in the linear regime the overall pattern of the torque density remains unchanged by varying the surface density slope, i.e. the interference of the modes is unchanged, only their individual amplitudes are rescaled with $\delta \Sigma_\mathrm{lin}$, see equation (\ref{eq:dslin}). See text for details.
}
\label{fig:dTdR_var_p}
\end{center}
\end{figure}


\section{Results: Variation of disc parameters in flux-preserving discs}
\label{sec:var_par}


Next we examine how the characteristics of the torque wiggles in the AMF-preserving discs are affected by variation of the disc parameters. Recall that $\dslin$ depends on $p$ and $q$ --- the slopes of the background surface density and temperature (which set the initial profile of entropy), see equations (\ref{eq:Sig_i})-(\ref{eq:T_i}) \& (\ref{eq:dslin}), while $\phil$ also depends on $\hp$, see equations (\ref{eq:csi_i}) \& (\ref{eq:phi-lin})-(\ref{eq:Deltaphi}). We will first explore in Section \ref{sec:wake_pars} how the self-similarity of the wake in the AMF-preserving discs is affected by the variation of $p$, $q$, and $\hp$, and then move on to study the effect of these parameters on the torque wiggles in Section \ref{sec:wiggles_pars}.


\subsection{Self-similarity of the wake structure}
\label{sec:wake_pars}


We illustrate the variation of the wake profile with disc parameters by examining the azimuthal dependence of the dimensionless profile shape $\psi(x)$ defined by the equation (\ref{eq:ansatz}) as a function of the (naturally scaled) azimuthal variable $(\phi - \phil)\hp^{-1}$. This allows us to eliminate the obvious dependencies of $\delta\Sigma$ on the disc parameters via $\dslin$ and $\phil$ and to focus on more subtle effects. These $\psi$-profiles are obtained via the linear calculation for several adiabatic disc models described below. 


\subsubsection{Variation of the slopes $p$ and $q$}
\label{sec:vary_slopes}

In Fig. \ref{fig:psi_var_pq} we show $\psi$ in the inner (left) and outer (right) disc with $\hp = 0.1$ for different values of $p$ and $q$ (colours) at several values of $R$ (columns). These plots reveal a pattern which is reminiscent of the discussion of the fiducial disc model in Section \ref{sec:planet_waves}. 

Namely, in the outer disc the wake profile remains roughly independent of $R$ suggestive of its approximate self-similarity. The variation of $p$ and $q$ introduces some deviations of $\psi(x)$ from its shape in the fiducial disc model (blue curve), but these differences are not significant. At the same time, in the inner disc we again observe the formation of additional maxima of $\psi$ which appear due to wake splitting into multiple spiral arms \citep{Bae17,Miranda2019I}. The structure of the inner wake does change noticeably as the disc parameters are varied, with the differences getting amplified towards the disc centre, e.g. see panel (c). It is also clear that the shape of the wake in the inner disc is affected mainly by the temperature slope $q$, which directly affects the emergence of additional spiral arms.

To summarize, the variation of $p$ and $q$ does not greatly modify the overall picture of wake propagation (for $\hp=0.1$): the outer wake maintains a self-similar shape with good accuracy, while the inner wake does not follow this universal pattern.


\subsubsection{Variation of $\hp$}
\label{sec:res_var_hp}


We now investigate the effect of varying $\hp$ on the wake shape, keeping $p=q=1$ fixed. In Fig. \ref{fig:psi_var_h} we show $\psi$ for several values of $\hp$, similar to Fig. \ref{fig:psi_var_pq}.
While the rescaling with $\dslin$ and stretching in $\phi$ by a factor $\propto \hp^{-1}$ allow us to compare the wake shape on similar scales in this plot, we see clear deviations between
the $\psi$ profiles for different $\hp$, even close to the planet (panels a \& d). 

As the distance from the planet increases, the wake suffers from the dispersive effects, which are especially pronounced for small $\hp$, consistent with the predictions of \citet{Miranda2019I}. This process is particularly noticeable in the inner disc (left), where the tertiary spiral becomes clearly visible at $R=0.1\Rp$ for $\hp=0.025$, while the $\hp=0.1$ disc shows only two clear spiral arms. 

What is more remarkable, for very low $\hp$ a secondary arm starts to develop even in the outer disc. This can be noticed by comparing the $\hp=0.025$ curves at $R=1.1\Rp$ and $3\Rp$ in the right columns of Fig. \ref{fig:psi_var_h}, revealing the development of a weak secondary arm far from the planet consistent with the analysis of \citet{Miranda2019I}. This indicates a modest violation of self-similarity within the radial region of interest occurring even in the outer disc for small enough $\hp$. 

To summarize, $\hp$ is a very important parameter affecting the density wave propagation more significantly than the changes in $p$ and $q$.


\subsection{Effect of disc parameters on torque wiggles}
\label{sec:wiggles_pars}


Next, we investigate the effect of changing disc parameters on the characteristics of torque wiggles. In Fig. \ref{fig:dTdR_summary} we show $\Texs$ computed both using full linear theory and using direct Athena++ simulations for a variety of disc parameters; a parameter that changes compared to the fiducial model (panel (a)) is labeled in each panel. Vertical magenta (dashed) lines correspond to the radial locations where the peak of the wake crosses the line $\phi=0$ from the star to the planet, while the green (dotted) lines correspond to the analytical prediction $\phil=2\pi n$, $n=1,2,...$; one can see that the two agree very well. Note that panels (a)-(d) all have $\hp = 0.1$. 

Comparing panels (a) and (b) we see that changing the surface density slope $p$ does not affect the radial periodicity of $\Texs$, i.e. the positions of prominent troughs (see also dashed vertical lines). This is expected since $\phil$ does not depend on $\Sigma_0$. However, the amplitude of the torque wiggles changes and in a non-uniform fashion: $p=0$ disc shows lower (higher) $\Texs$ in the inner (outer) disc compared to $p=1$. This can be understood from the dependence of the excitation torque density on $\Sigma_0(R)$: $\Texs\propto \delta\Sigma \propto \dslin\propto \Sigma_0^{1/2}(R)$, see equation (\ref{eq:dT}). Thus, given their approximately identical  $\psi(x)$ for discs with different $p$ (see Fig. \ref{fig:psi_var_pq}d-f), we expect that curves of $\Sigma_0^{-1/2}\Texs$ would have a universal shape. This is indeed the case, as illustrated by Fig. \ref{fig:dTdR_var_p} for $p=1$ (solid) and $p=0$ (dashed): with this normalization the two curves track each other closely over the entire radial range.

Panels (c) and (d) illustrate torque wiggles in discs with different initial entropy profiles by considering a reduced temperature slope $q = 0.5$ and a constant temperature ($q = 0$).  It is clear that the variation of $q$ leads to changes of not only the amplitude of the wiggles and but also their radial periodicity --- the distance between the consecutive troughs increases (decreases) in the outer (inner) disc as $q$ goes down.  This is explained via the behaviour of $\phil$, which determines the global shape of the wake.  Regarding the amplitude, equation (\ref{eq:dslin}) predicts $\Texs \propto \dslin \propto (R/\Rp)^{3q/4}$ in AMF-preserving discs, such that lower $q$ should increase (decrease) $\Texs$ in the inner (outer) disc. This is consistent with the trends seen in panels (c) and (d).

Finally, in panels (e), (f) we consider disc models with lower $\hp$,  while keeping $p=q=1$ as in panel (a). Comparing the locations of wake crossings with the radial structure of $\Texs$ it is evident that when $\hp$ is halved, the number of wake-crossings doubles, as expected from $\phil \propto \hp^{-1}$.  Closer examination of  panel (e) reveals that in the outer disc with $\hp =0.05$, wake-crossings are again aligned with minima of $\Texs$ for the first few radial periods. However, these minima are preceded by a maximum of similar magnitude, contrary to the fiducial case, where the troughs clearly dominate. This indicates that the variation of the shape of $\psi$ as $\hp$ is decreased (see Fig. \ref{fig:psi_var_h}) has a direct effect on $\Texs$. For even lower $\hp = 0.025$ (panel f), the wake-crossings in the outer disc align with the dominant peak or trough of $\Texs$ even less; one also notices variation of the wiggle pattern, even between neighbouring (e.g. first and second) wake-crossings. Again, this is clearly caused by the erosion of the wake self-similarity for low $\hp$, see Section \ref{sec:res_var_hp}.

The amplitude of variations in $\Texs$ (normalized by $F_{J,0}$ at fixed $R$) in panels (e) and (f) also clearly decreases with $\hp$, compatible with the expectation $\Texs \times F_{J,0}^{-1} \propto \dslin  \hp^{3} \propto \hp^{3/2}$, see equations (\ref{eq:FJ0}) and (\ref{eq:dslin}).


\section{Torque wiggles in non-AMF-conserving discs}
\label{sec:res_beta}


In this section we study the effect of relaxing the wave AMF conservation on the existence and properties of torque wiggles. As an example of a non-AMF-conserving disc we will consider a disc with thermal relaxation in the form of $\beta$-cooling, see Section \ref{sec:disc_model} and Appendix \ref{sec:beta}. The propagation of planet-driven waves in such discs has been previously explored by \citet{Miranda2020I}. We illustrate our results using several values of $\beta$ ranging from $10^{-3}$ to $10^2$, which allows us to explore the transition from an isothermal to adiabatic thermodynamics. We also consider the pure locally isothermal limit $\beta\to 0$ \citep{Miranda2019II}.


\begin{figure*}
    \begin{center}
    \includegraphics[width=0.99\textwidth]{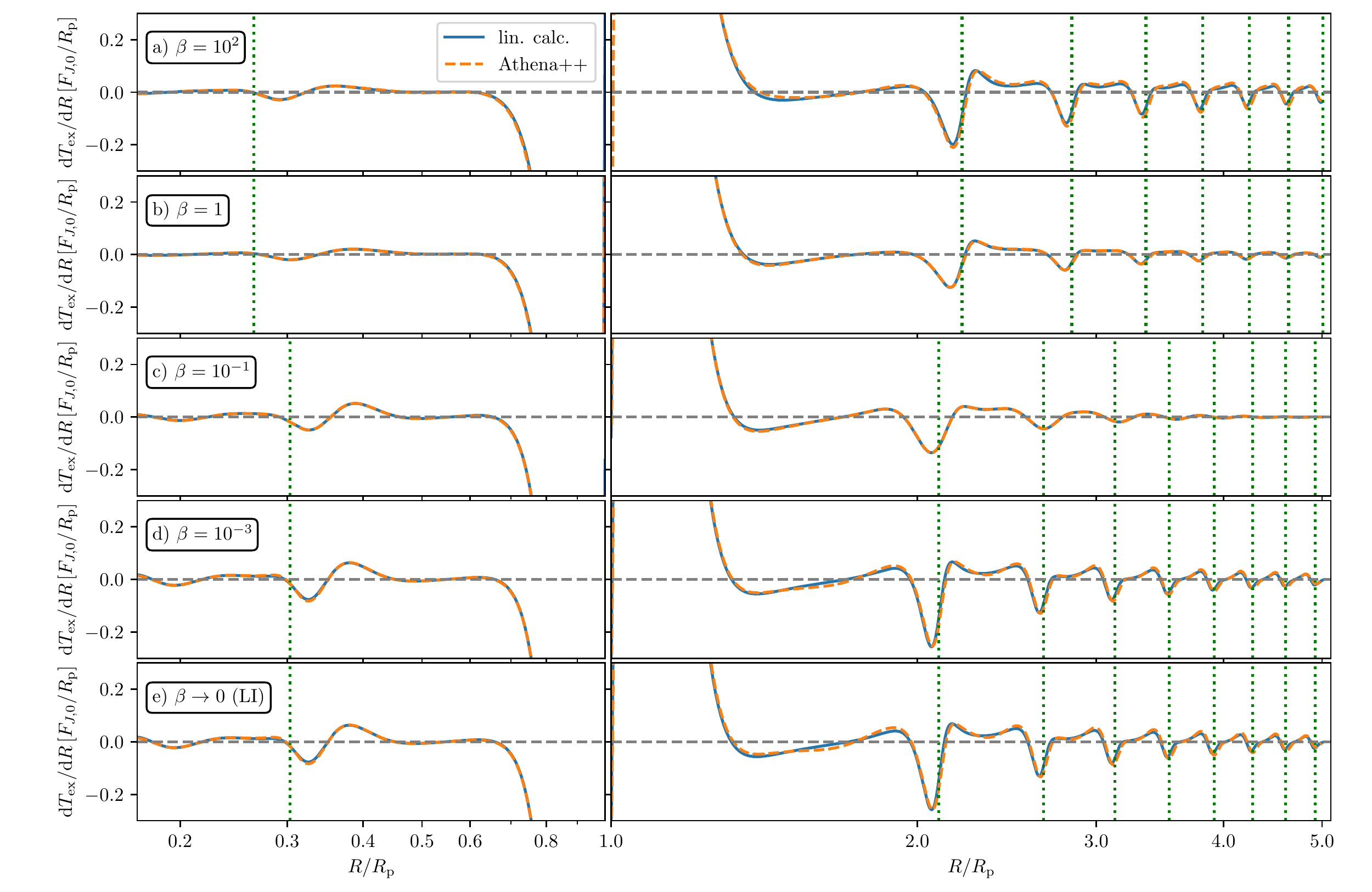}
    \caption{
    Same as Fig. \ref{fig:dTdR_summary} but now illustrating the effect of varying the dimensionless cooling time $\beta$ (decreasing from top to bottom) in calculations with $\beta$-cooling (for the fiducial values $p=q=1$, $\hp = 0.1$). Calculation of the locations of the vertical green dotted lines ($\phil(R)=2\pi n$, $n=1,2,...$) uses $\csa$ in panels (a),(b) and $\csi$ in all others (see Section \ref{sec:disc_model}). See text for a discussion.
    }
    \label{fig:dTdR_betas}
    \end{center}
\end{figure*}



\begin{figure}
    \begin{center}
    \includegraphics[width=0.49\textwidth]{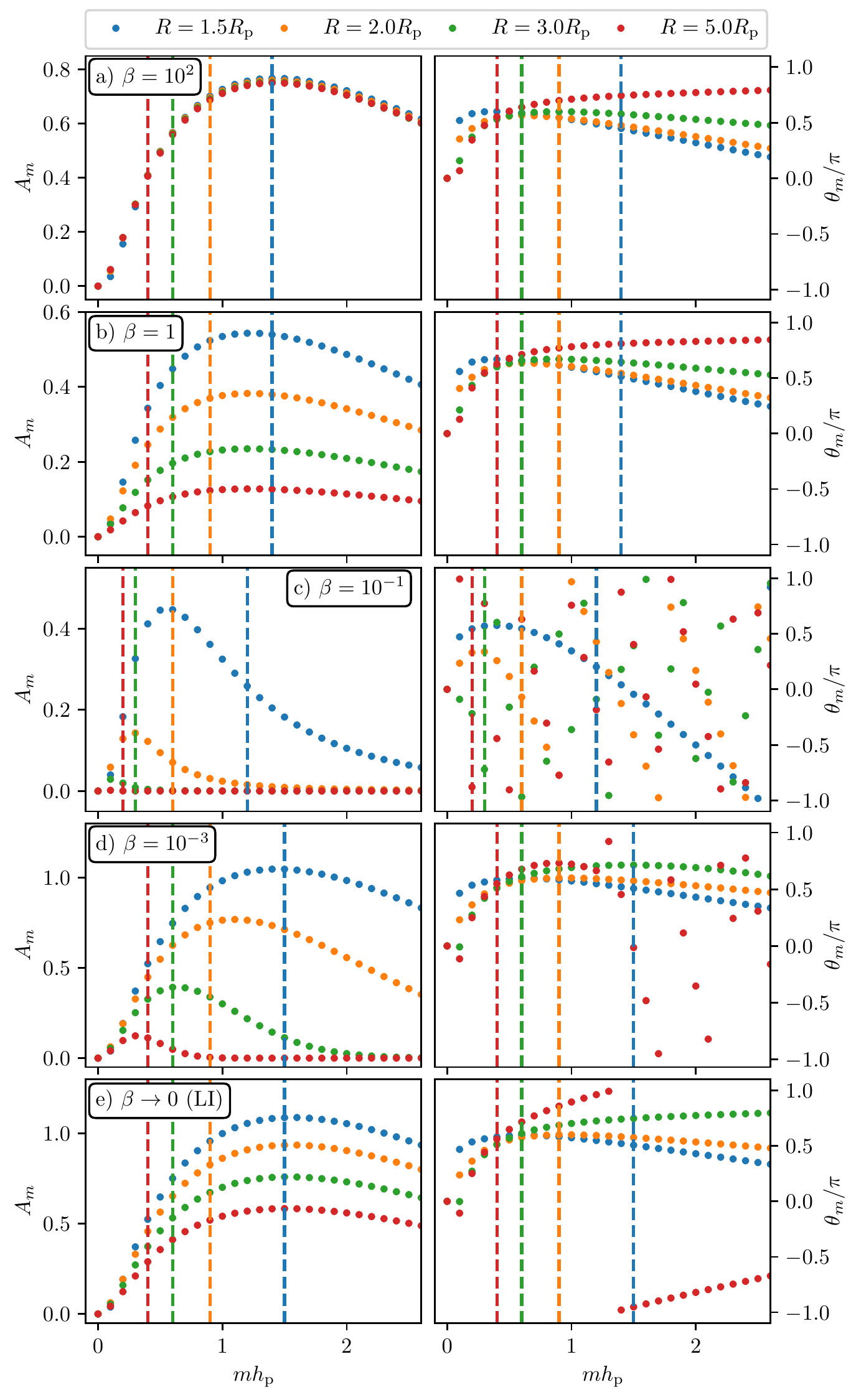}
    \vspace*{-0.5cm}
    \caption{
        Amplitudes $A_m$ and phases $\theta_m$ of $\Psi_m$ (similar to Fig. \ref{fig:amp_phase_fid}) sampled at different $R>\Rp$ (shown at the top) for disc models with $\beta$-cooling and the fiducial disc parameters $p=q=1$, $\hp = 0.1$. The cooling time $\beta$ (indicated in each row) takes the same values as in Fig. \ref{fig:dTdR_betas}. The vertical dashed lines have the same meaning as in Fig. \ref{fig:amp_phase_fid}.
    }
    \label{fig:amp_phase_var_beta}
    \end{center}
\end{figure}


We show $\Texs$ results for different values of the dimensionless cooling time $\beta$ (with the fiducial set of disc parameters) in Fig. \ref{fig:dTdR_betas}, again using both the linear calculation and simulations (which are in excellent agreement). A first glance reveals that the variation of $\beta$ clearly has an effect on the amplitude and radial periodicity of the torque wiggles, both in the inner and outer disc. 

As $\beta$ decreases, the radial pattern of repeating troughs shifts closer to the planet. This can be understood as follows: for $\beta\gg 1$ the thermodynamic response is approximately adiabatic and $\cs\approx \csa = \gamma^{1/2} \csi$ --- i.e. the adiabatic sound speed, which because of the factor $\gamma>1$ is higher than the isothermal sound speed $\csi$, to which $\cs$ reduces in the opposite limit $\beta\ll 1$. As a result, equation (\ref{eq:phi-lin}) predicts slower evolution of $\phil$ with $R$, i.e. a less tightly wound density wave for $\beta\gtrsim 1$ as compared to $\beta\lesssim 1$ case \citep[see also][]{Miranda2020I}. This point is illustrated using vertical blue (dotted) lines corresponding to $\phil=2\pi n$, $n=1,2,...$, which are computed using $\cs\approx \csa$ in  (\ref{eq:phi-lin}) for $\beta\ge 1$ and using $\cs\approx \csi$ for $\beta< 1$; these lines are well aligned with the troughs of $\Texs$. 

The amplitudes of torque wiggles are quite similar in the adiabatic ($\beta=10^2$) and locally isothermal limits, compare panels (a) and (e). At the same time, the wiggles are strongly suppressed for intermediate values of $\beta$, especially in $\beta=0.1$ disc (panel c) and at large $R$. These trends can be easily understood by examining the behaviour of the amplitudes $A_m$ and phases $\theta_m$ of the Fourier coefficients $\Psi_m$ (see Section \ref{sec:modal}) in the outer disc, which are shown in Fig. \ref{fig:amp_phase_var_beta}. In doing this it is helpful to keep track of the vertical colored lines at $m_\mathrm{c}$ --- the maximum $m$ significantly contributing to $\Texs$ --- at every $R$, see Fig. \ref{fig:amp_phase_fid} and Section \ref{sec:modal}.

For $\beta=10^2$ (panel (a)), close to adiabatic limit, $A_m$ and $\theta_m$ behave as they do in Fig. \ref{fig:amp_phase_fid} --- the former are independent of $R$, while the latter slowly evolve with $R$. As a result, the wiggle pattern is essentially indistinguishable from that of Fig. \ref{fig:dTdR_summary}a. For $\beta=1$ we see $A_m$ being uniformly (in $m$) suppressed as $R$ increases, which is caused by the decay of the wave AMF due to thermal relaxation for intermediate values of $\beta$, as described in \citet{Miranda2020I}. In particular, note the obvious reduction of $A_m$ to the left of the corresponding $m=m_\mathrm{c}$ lines. This reduces the amplitude of the wiggles seen in Fig. \ref{fig:dTdR_betas}b. 

Wave decay is most dramatic for $\beta = 0.1$ (Fig. \ref{fig:amp_phase_var_beta}c), leading to the precipitous drop of $A_m$ for $m<m_\mathrm{c}$  and the corresponding rapid decay of the torque wiggle amplitude with $R$ in Fig. \ref{fig:dTdR_betas}c. Moreover, for this value of $\beta$ the phases $\theta_m$ also show rather erratic behaviour, indicative of the loss of constructive interference of different Fourier harmonics, again consistent with \citet{Miranda2020I} findings.

The recovery of the wiggle amplitude for $\beta=10^{-3}$, see Fig. \ref{fig:dTdR_betas}d, is due to the disc starting to transition to the locally isothermal regime. Fig. \ref{fig:amp_phase_var_beta}d shows that $A_m$ decays with $R$ slower than in panel (c), in particular, the most significant $A_m(m)$ (for $m<m_\mathrm{c}$) for  every $R$ are not that different even from panel (a). This agreement is improved even further in the locally isothermal limit shown in Fig. \ref{fig:amp_phase_var_beta}e (although here the suppression of $A_m$ with $R$ is more uniform\footnote{\citet{Miranda2020I} have shown that in the locally isothermal discs the angular momentum carried by $m$-th azimuthal harmonic of the wave scales as $F_{J,m}\propto \cs^2(R)$, independent of $m$.} in $m$, more akin to panel (b)), explaining the qualitative similarity of $\Texs$ in panels (a) and (e) of Fig. \ref{fig:dTdR_betas} (modulo the differences in the radial periodicity).

To summarize, the evolution of the torque wiggles with the cooling time in discs with $\beta$-cooling is entirely consistent with the results of \citet{Miranda2020I} on the damping of planet-driven density waves in such discs.


\section{Discussion}
\label{sec:disc}


In this work we provided a detailed exploration of the properties and origin of the torque wiggles --- a new feature of the global disc-planet interaction which has been seen in only a handful of studies so far \citep{AR18P,Miranda2019I,Miranda2019II,Dempsey2020}. Together with the 'negative torque density phenomenon' discovered in the local limit \citep{RP12}, this feature represents an interesting departure from the simple $\Texs\propto\vert R-\Rp\vert^{-4}$ picture that has been prevalent in the field for decades  \citep[e.g.][]{Armitage2002,Chang2010,Zagaria2021}, and paves a way to understanding some key aspects of the tidal disc-perturber coupling in other settings. 

Our study clearly shows that torque wiggles owe their existence to the {\it global} structure of the planet-driven density wave and its direct coupling to the planetary gravity. Periodicity of the wiggles arises because of the wave wrapping due to the differential rotation, making possible azimuthal alignments of the wake with the planet, which result in sharp features of $\Texs$ (Section \ref{sec:asy_analysis}). This situation would not be possible in the (infinitely extended) shearing sheet geometry, meaning that torque wiggles do not arise in the local limit\footnote{A shearing-box setup with periodic boundary conditions in the pseudo-azimuthal direction that extends far enough in the pseudo-radial direction might allow for similar features.}.
This is consistent with the calculations of \citet{RP12} who showed that the negative torque density phenomenon (i.e. no additional $\Texs$ reversals) is the only possibility far from the planet in the shearing sheet. It is important to emphasize that the torque wiggles are not related to the low-order Lindblad resonances \citep{GT80} which are not present far from the planet.

In a fully global setting, we have shown the torque wiggles to be a very robust and ubiquitous phenomenon. They are present for different disc parameters (Section \ref{sec:var_par}), different assumptions about the disc thermodynamics (Section \ref{sec:res_beta}), with and without accounting for the non-linear effects (more on this in Section \ref{sec:lin_vs_nonlin}). Torque wiggles were originally seen in 3D simulations of \citet{AR18P}, but we clearly see them also in the 2D setting. Also, the work of \citet{AR18P} accounts for the indirect potential\footnote{We again stress that \citet{AR18P} and other studies, including this work, calculate the torque on the disc accounting for only the direct gravitational force of the planet (i.e. do not consider the indirect force when computing the excitation torque).} when computing the response of the disc to the planetary perturbation, whereas our study (as well as \citealt{Miranda2019I,Miranda2019II} and many other works) neglects the indirect potential. This implies that the dimensionality of the problem and indirect potential are not the key factors for the origin of the torque wiggles. 

Self-similarity of the wake (even if approximate) is another important ingredient for the appearance of a regular pattern of torque wiggles. Rapid dispersive decoherence of the wake in the inner disc resulting in the wake splitting into multiple spiral arms is one of the reason why the wiggles are aperiodic and largely suppressed for $R<\Rp$ (another one is the faster radial decay of $\Texs$ there, see Section \ref{sec:asy_analysis}). In the outer disc this decoherence is much weaker as shown by \citet{Miranda2019I} and the wake maintains a self-similar form with good accuracy (except for very low $\hp$). We note that in the presence of a well-developed turbulence, the wake coherence may be destroyed even in the outer disc \citep{Zhu2013}, resulting in the suppression of the torque wiggles (although they may still be evident after time-averaging).

Shear viscosity in a (weakly-turbulent) hydrodynamic disc is unlikely to challenge the overall picture of the torque wiggles. First, \citet{Miranda2020II} have shown that viscous dissipation is ineffective at damping the density waves as compared to other processes --- radiative and non-linear damping. Thus, the wake is likely to maintain its global structure unless the viscosity is very large, $\alpha\gtrsim 0.1$. Second, viscous dissipation of the wake would mainly damp the high-$m$ Fourier modes contributing to the wake. However, we have seen in Section \ref{sec:rec_tex_fid} that it is mainly the low-$m$ modes that dominate in the genesis of the torque wiggles, and increasingly so at large $R$. Our own simulations (which we do not show here) demonstrate that even for effective viscosity as high as $\alpha \leq 10^{-2}$ the appearance of torque wiggles does not change much.


\subsection{Effects of wave non-linearity}
\label{sec:lin_vs_nonlin}


\begin{figure}
    \begin{center}
    \includegraphics[width=0.49\textwidth]{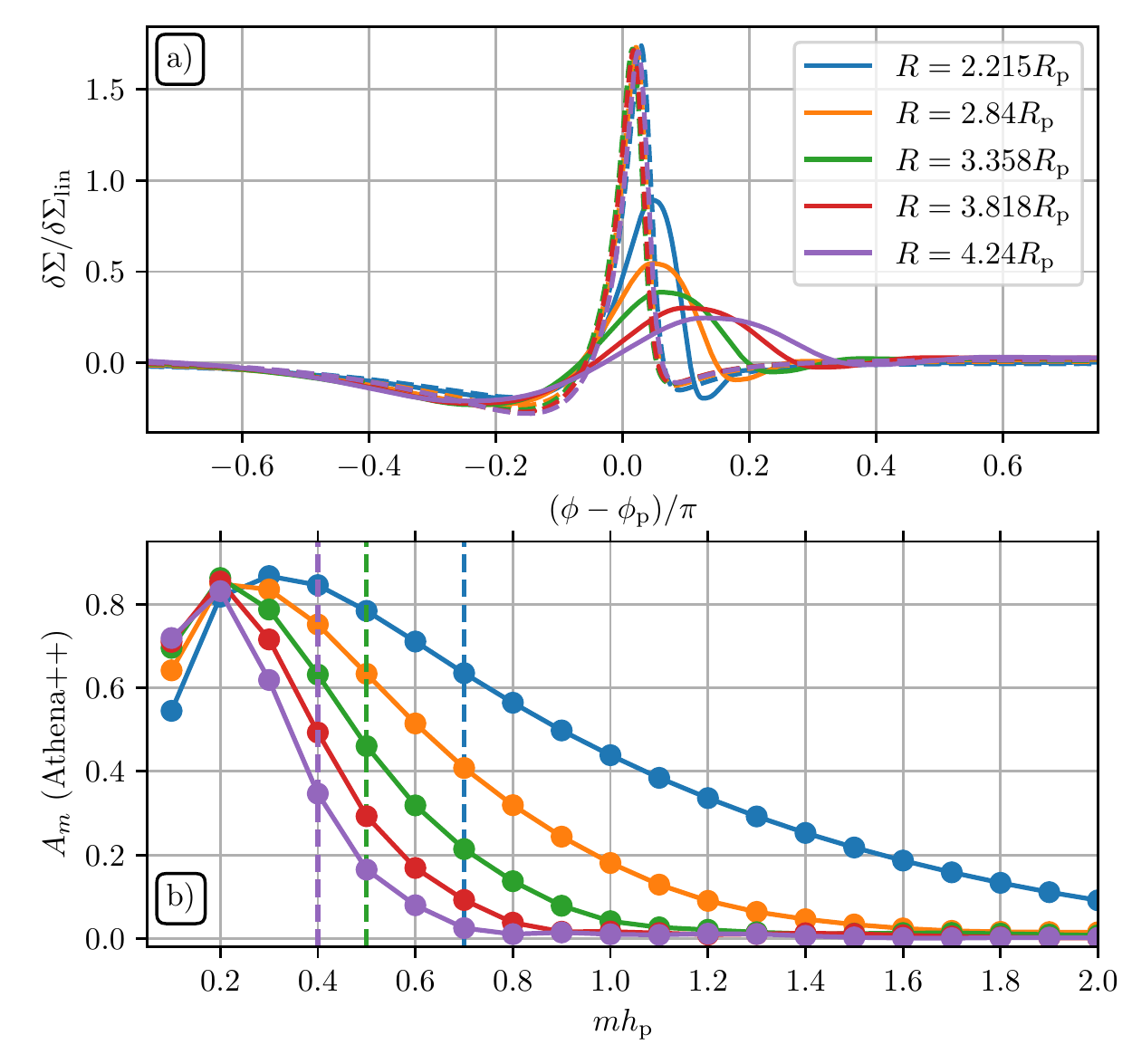}
    \vspace*{-.5cm}
    \caption{
    {\it Top}: Azimuthal profiles of the relative surface density perturbation obtained using Athena++ simulations with fiducial parameters and $\Mp=0.01\Mth$ (solid) and linear calculations (dashed, falling almost on top of each other) at locations where $\phi_\mathrm{lin} = \phip$. Even for a low $\Mp$, non-linearity leads to substantial deviations from linear theory at large distances from the planet (see also \citet{Cimerman2021}).
    {\it Bottom:} Amplitude $A_m$ of Fourier components of $\Psi_m$ for the non-linear Athena++ simulations as a function of wave-number $m$, measured at the same radii as in the top panel; vertical dashed mark $m_\mathrm{c}$. As opposed to the linear solution (cf. Fig. \ref{fig:amp_phase_fid}b), there is a clear evolution of $A_m(m)$ with $R$ leading to efficient transfer of power to low $m=1,2$ modes, as well as the overall decay at high $m$ due to shock-damping.
    }
    \label{fig:fid_sigma_r}
    \end{center}
    \end{figure}

\begin{figure*}
    \begin{center}
    \includegraphics[width=0.99\textwidth]{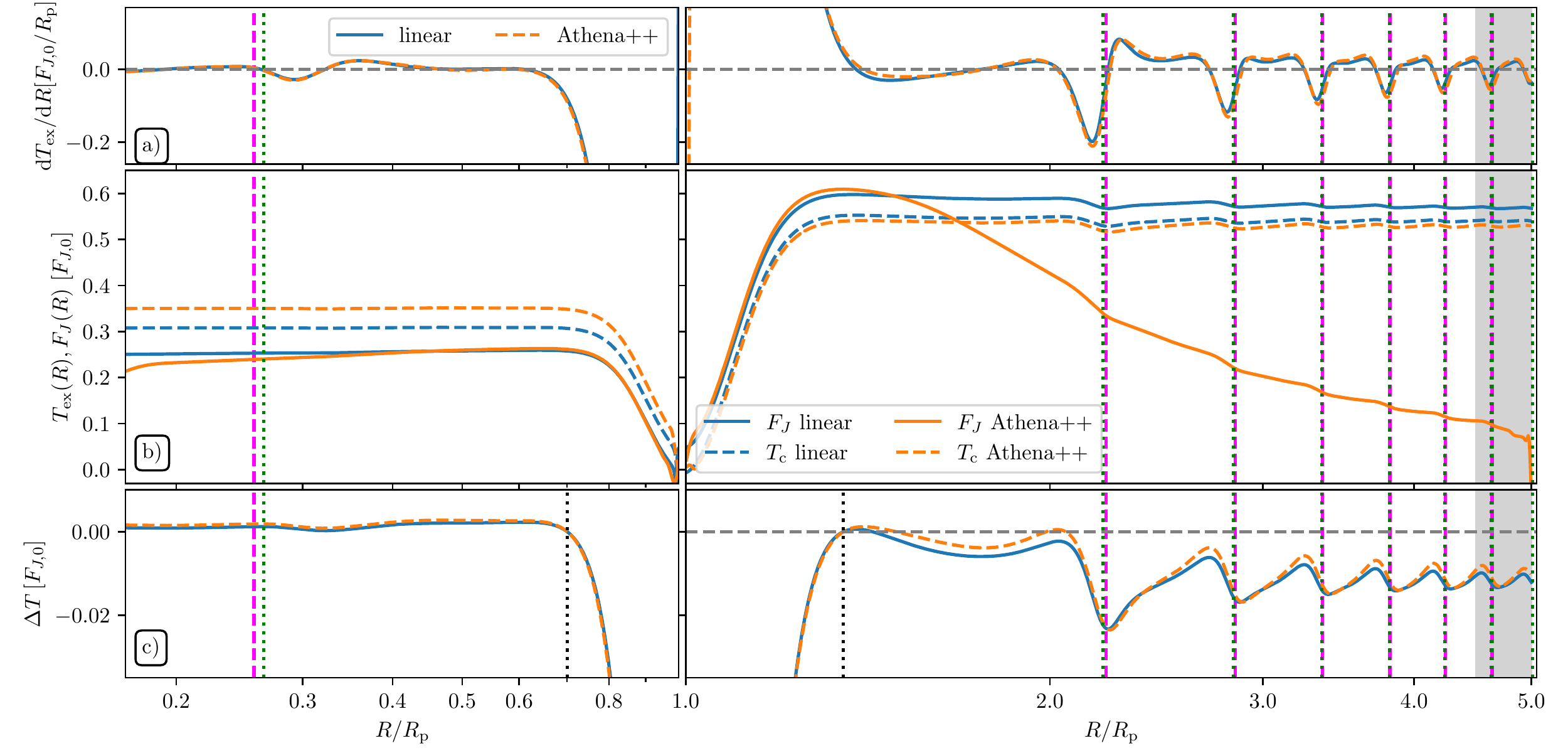}
    \vspace*{-.1cm}
    \caption{
    (a) Radial excitation torque density $\Texs$ (same data as in Fig. \ref{fig:dTdR_summary}a. (b) Integrated (total) torques $T_\mathrm{ex}(R)$ (dashed) and AMF $F_J(R)$ (solid) obtained using linear theory (blue) and simulations (orange). (c) Global variations in the total torque relative to its value at some reference radius $R_\mathrm{r}$, see text for details. As before, the vertical lines mark radii where either $\phil = \phip$ (green) and at the position of the peak of $\delta \Sigma$, $\phipk = \phip$ (purple). See Section \ref{sec:int_tq_fid} for details. 
    }
    \label{fig:fid_dtdr_int_fj}
    \end{center}
\end{figure*}

Nonlinear effects are important for propagation of the planet-driven density wave: the wake steepens into a shock, which leads to wave dissipation. This process is inevitable even for low-mass planets, see Fig. \ref{fig:fid_sigma_r}a in which we illustrate the non-linear evolution of the azimuthal profile of the wave in a fiducial disc model for $\Mp=0.01\Mth$. While the profiles computed using linear theory (dashed curves) show almost no change with $R$, $\delta\Sigma$ derived from simulations fully accounting for the wave non-linearity evolve significantly: as $R$ increases, the wave stretches azimuthally and its amplitude goes down. This is reflected also in the behaviour of the angular momentum flux carried by the wave as illustrated in Fig. \ref{fig:fid_dtdr_int_fj}, where in panel (b) the orange curve shows a clear decay\footnote{The decay of $F_J$ is far less dramatic in the inner disc because (a) for this disc model the nonlinear evolution is faster in the outer disc (see Fig. 2 of \citealt{Cimerman2021}) and (b) formation of a secondary spiral arm which reduces the amplitude of the primary arm, slowing down its nonlinear evolution.} of $F_J$ (extracted from our simulations) in the outer disc after the wave has shocked. One would expect this wave damping to have a direct impact on the appearance of torque wiggles. 

However, comparison of $\Texs$ profiles computed using linear theory and simulations in Figs. \ref{fig:dTdR_summary} \& \ref{fig:dTdR_betas} shows that this is not the case and, at least for the low $\Mp$ used in simulations, nonlinear damping of the wave has rather little impact on the torque wiggles. This unexpected outcome can be understood based on the fact that far from the planet $\Texs$ behaviour is determined only by a handful of low-$m$ Fourier harmonics of $\delta\Sigma$, see Section \ref{sec:rec_tex_fid} and Fig. \ref{fig:conv_series_fid}. Moreover, Fig. \ref{fig:fid_sigma_r}b shows that while the amplitudes (represented by $A_m$) of high-$m$ harmonics rapidly die out, the amplitudes of the low $m=1,2$ harmonics tend to {\it grow} with $R$, which happens because of the nonlinear azimuthal stretching of the profile, see panel (a). These two effects conspire to keep the amplitudes of the torque wiggles essentially the same for $\Mp=0.01\Mth$, despite the nonlinear damping of the wake.  

We expect that for higher $\Mp$ the nonlinear wave damping will have a more dramatic effect on the torque wiggles, reducing their amplitude; this is supported by our preliminary runs in high $\Mp$ regime which we do not show here.  Nevertheless, Fig. 1 of \citet{Miranda2019II} still shows some torque wiggles up to $\Mp=\Mth$, see also \citet{Dempsey2020}.  When wave nonlinearity is strong,  the shape of the torque wiggles may become sensitive to the slope of the surface density $p$ (unlike in the linear case, see Fig. \ref{fig:dTdR_var_p}), since the rate of nonlinear evolution of the wave depends on it  \citep{R02,Cimerman2021}.


\subsection{Effect of torque wiggles on the total torque}
\label{sec:int_tq_fid}


Fig. \ref{fig:fid_dtdr_int_fj} also illustrates the impact of the torque wiggles on the full (integrated) one-sided torque exerted by the planet on the disc, by showing the behaviour of the cumulative torque (dashed curves in panel (b))
\begin{align}
    T_\mathrm{ex}(R) = \int_{\Rp}^{R} \Tex \,\de R.
\end{align}

One can see that close to the planet $T_\mathrm{ex}(R)$ follows $F_J(R)$ quite closely, except for the small vertical offset related to the torque in the corotation region. The two start to diverge after the wave shocks and starts transferring its angular momentum to the disc fluid, with AMF decaying below $T_\mathrm{ex}(R)$ in the nonlinear case. Despite that, both $T_\mathrm{ex}(R)$ and $F_J(R)$ clearly show the low-amplitude oscillations caused by the torque wiggles. It is interesting that the torque wiggles have an impact (albeit small) on the wave AMF even far from the planet, where the wave is usually considered to be freely propagating and no longer affected by the planetary potential. 

These oscillations are more easily seen in panel (c), where we plot $\Delta T = T_\mathrm{ex}(R) - T_\mathrm{ex}(R_\mathrm{r})$ --- the difference between the cumulative torque at $R$ and the cumulative torque at some reference $R_\mathrm{r}$. We pick somewhat arbitrarily $R_\mathrm{r,i} = 0.7 \Rp$ and $R_\mathrm{r,o} = 1.3 \Rp$ in the inner and outer disc, respectively (dotted black vertical lines), just outside the main peaks of $\Texs$. In the inner disc, the total torque hardly changes and $\Delta T$ remains close to zero for $R<R_\mathrm{r,i}$. In the outer disc, however, we clearly see the variations due to the integrated effect of the torque wiggles. The greatest deviation from the maximum $T_\mathrm{ex}$ is $\Delta T/T \simeq -0.03$ and its location coincides with the first wake crossing of $\phip$ around $R \simeq 2.2 \Rp$. These oscillations eventually die out as $R\to \infty$, however their integrated effect is non-zero and $\Delta T(\infty)/T \simeq -0.015$, see panels (b) and (c). Thus, the integrated effect of the torque wiggles in the outer disc is to slightly {\it lower} $T_\mathrm{ex}$ from the maximum value near $R\approx 1.3\Rp$.

In our models, the highest relative variation of $T_\mathrm{ex}$ in the outer disc due to torque wiggles is $\Delta T/ T_\mathrm{ex} \simeq 0.07$ (after the first wake-crossing) for $p=0$, $q=0$, $\hp=0.1$ disc. Clearly this effect is small, as is the amplitude of torque wiggles in general, as the main peak of $\Texs$ within a couple of scaleheights from the planet provides the dominant contribution to $T_\mathrm{ex}$, see Fig. \ref{fig:dtdr_overview}. Moreover, $T_\mathrm{ex}$ is the integral of the radially oscillating $\Texs$ see equation (\ref{eq:dT4}), which additionally lowers $\Delta T$. However, things would be different if the planet were more massive and opened a gap around its orbit. The reduced surface density would suppress the near-planet peak of $\Texs$, accentuating the effects of the torque wiggles, see Fig. 5 of \citet{Dempsey2020} for an illustration of this trend. A study of the torque behaviour in this regime will be presented in Cimerman \& Rafikov (in prep.).


\section{Summary}
\label{sec:disc_conclusion}


In this work we explored the physics of torque wiggles --- the conspicuous low-amplitude features in the radial profile of the torque density (due to the direct force from the planet) at large distances ($\vert R-\Rp\vert\gtrsim \Rp$) from a planet that launches a density wave in a protoplanetary disc. These features are characterized by a remarkable pattern of radial variability (Section \ref{sec:heuristic}) and have been previously seen in a number of simulations. We explored their properties using linear theory and direct hydrodynamic simulations (finding good agreement between them in appropriate limit of low planet masses), and developed a theory explaining their origin. We summarize our main findings below.

\begin{itemize}

\item The wiggles arise due to the global gravitational coupling of the planetary potential to the planet-driven density wave propagating through the disc. They appear predominantly in the outer disc, where their amplitude is substantial and their radial periodicity is obvious (Section \ref{sec:outer_typical}). They are suppressed and are irregular in the inner disc (Section \ref{sec:inner_typical}). The following statements are for the outer disc. 

\item The wiggles appear for all disc parameters that we explored in this work (Section \ref{sec:var_par}) and for various assumptions about the disc thermodynamics, although they are strongly suppressed in discs with (the dimensionless cooling time) $\beta\sim 1$ (Section \ref{sec:res_beta}). They appear in both 2D and 3D discs \citep{AR18P}.

\item While we mainly explore the wiggle properties using linear theory, they are also remarkably insensitive to the non-linear effects (Section \ref{sec:lin_vs_nonlin}). Their amplitude scales as $\Mp^2$, just as that of the full planetary torque.

\item We developed analytical theory (Section \ref{sec:theory}) showing that quasi-periodic wiggles arise because of the global nature of the planet-driven density wave (its multiple wrappings in the disc) and the approximately self-similar behaviour (Section \ref{sec:planet_waves}) of the wave in the outer disc (which starts getting violated as $\hp\to 0$).

\item This theory demonstrates that the behaviour of $\Texs$ at $\vert R-\Rp\vert\gtrsim \Rp$ is dominated with good accuracy by a small number of low-$m$ Fourier components of the perturbation $\delta\Sigma$ (Section \ref{sec:rec_tex_fid}). 

\item This theory also explains the key features of $\Texs$ behaviour (Section \ref{sec:asy_analysis}). In particular, it accounts for the sharp features in $\Texs$ that appear at the radii where the planetary wake crosses the star-planet line as a result of constructive interference of the most significant low-$m$ modes of the density perturbation (in the inner disc violation of the constructive interference related to the formation of multiple spiral arms suppresses the wiggles). 

\item While the amplitude of the wiggles is low in the linear regime and they affect the total (one-sided) torque imparted by the planet on the disc only at the level of several per cent (Section \ref{sec:int_tq_fid}), their significance will grow for higher planet masses when the gap opening becomes important.

\end{itemize}

Future work (Cimerman \& Rafikov, in prep.) will explore torque behaviour in the non-linear regime of circumbinary discs, when the perturber-to-star mass ratio $q\sim 1$. We will demonstrate that the torque wiggles become much more important in this regime.


\section*{Acknowledgements}


We are grateful to Lev Arzamasskiy for useful discussions. N.P.C. is funded by an Isaac Newton Studentship and a Science and Technology Facilities Council (STFC) studentship. R.R.R. acknowledges financial support through the Ambrose Monell Foundation, and STFC grant ST/T00049X/1. A large part of the long term simulations were performed on the HPC cluster FAWCETT at DAMTP, University of Cambridge. Part of this work was performed using resources provided by the Cambridge Service for Data Driven Discovery (CSD3) operated by the University of Cambridge Research Computing Service (\texttt{www.csd3.cam.ac.uk}), provided by Dell EMC and Intel using Tier-2 funding from the Engineering and Physical Sciences Research Council (capital grant EP/P020259/1), and DiRAC funding from the Science and Technology Facilities Council (\texttt{www.dirac.ac.uk}).

\noindent\textit{Software:} NumPy \citep{2020NumPy-Array}, SciPy \citep{2020SciPy-NMeth}, IPython \citep{IPython}, Matplotlib \citep{Matplotlib}, Athena++ \citep{Athenapp2020}.

\section*{Data Availability}
The data underlying this article will be shared on reasonable request to the corresponding author.




\bibliographystyle{mnras}
\bibliography{references} 




\appendix


\section{Numerical methods}
\label{sec:num_methods}


\subsection{Setup of the linear solver}
\label{sec:lin_solv}


To obtain linear solution for the planet-induced perturbation we employ the exact same machinery as in \citet{Miranda2019I,Miranda2020I}. The only difference is that now we use a more accurate approximation for the planetary potential with a softening prescription that converges to a Newtonian potential at higher (fourth) order close to the planet \citep{Dong2011}:
\begin{align}
    \Phi_\mathrm{p}=\Phi_\mathrm{p}^{(4)} = -G M_\mathrm{p} \frac{d^2 + (3/2) r_\mathrm{s}^2}{\left(d^2 + r_\mathrm{s}^2 \right)^{3/2}},
    \label{eq:phi4_p2}
\end{align}
where $d = \vert \vect{r} - \vect{r}_\mathrm{p} \vert$ is the distance from the planet and $r_\mathrm{s} = \epsilon H(R=R_\mathrm{p})$ is the smoothing length for which we adopt $\epsilon = 0.6$. This choice is consistent with our previous work \citep[][see Appendix A therein]{Cimerman2021}.

The radial mode solutions are obtained on a grid ranging from $R_\mathrm{in} = 0.05 \Rp$ to
$R_\mathrm{out}= 5 \Rp$ consisting of $N^\mathrm{lin}_R = 10^5$ logarithmically spaced cells for all values of $\hp > 0.025$. For this lowest value, we consider a smaller radial domain ranging from $R_\mathrm{in} = 0.1 \Rp$ to $R_\mathrm{out}= 3 \Rp$ and keep $N^\mathrm{lin}_R$ fixed to increase the radial resolution, which
is needed as the wave pattern becomes very tightly wound.

The number of modes $\mmax$ that are typically needed for a converged solution depends on the disc scale-height $\hp$. In particular, we used $\mmax = 100, 220, 320$ for $\hp = 0.1, 0.05, 0.025$ respectively. These values were chosen on a case by case basis until convergence was found.\footnote{For $\hp = 0.1$ and $\hp = 0.05$, we included more modes than needed, since they were easy enough to obtain.}


\subsection{Setup of non-linear simulations}
\label{sec:setup_nl}


Our setup is similar to the one used in \citet{Cimerman2021}, however here we additionally solve an equation for the total energy of the fluid, to allow for initial temperature gradients (setting the initial entropy) in the disc. The total set of equations solved is
\begin{align}
    \pd{\rho}{t} + \nabla \cdot (\rho \vect{u}) &= 0, \\
    \pd{(\rho \vect{u})}{t} + \nabla \cdot (\rho \vect{u} \otimes \vect{u} + P \vect{I}) &= - \rho \nabla \Phi,\\
    \pd{E}{t} + \nabla \cdot \left[(E + P)\vect{u} \right] &= - \rho \vect{u} \cdot \nabla \Phi + \Lambda,
\end{align}
where $E = \epsilon + \rho\abs{\vect{u}}^2/2$ is the total energy density, $\epsilon = e \rho$ the internal energy density, $P = (\gamma-1) \epsilon$ the gas pressure with the ratio of specific heats $\gamma$, $\vect{I}$ the identity tensor,
$\Phi = \Phi_\star + \Phi_\mathrm{p}$ is the total gravitational potential , where $\Phi_\star = -GM_\star/R$, and $\Phi_\mathrm{p}$ (given by equation \ref{eq:phi4_p2}), are due to the central star and the planet, respectively (i.e. no indirect potential). The source term $\Lambda$ allows for thermal relaxation to a fixed background and is described in Appendix \ref{sec:beta}.

The simulation domain ranges over $0.1 \leq R/\Rp \leq 5.0.$ in radius and is $0 \leq \phi \leq 2\pi$ in azimuth. We use $N_R \times N_\phi = 2240 \times 3600 $ cells in the fiducial setup (when $\hp = 0.1$) that are logarithmically spaced in radius and uniformly spaced in azimuth. For cases with $\hp = 0.05$, this resolution is doubled. We do not perform an Athena++ simulation for $\hp = 0.025$ due to the excessive resolution requirements. We have checked that our main findings are not influenced by resolution by performing a double resolution simulation for the fiducial model.

Wave-damping zones \citep{dvB2007II} are implemented close to the radial boundaries in the zones $0.1 \leq R/\Rp \leq 0.15$ and $4.5 \leq R/\Rp \leq 5$ in order to avoid reflections
\citep[see][]{Cimerman2021}.


\section{$\beta$-cooling prescription}
\label{sec:beta}


Following \citet{Miranda2020I}, we use a prescription where thermal relaxation towards the unperturbed specific internal energy density $e = \epsilon/\Sigma = \csa^2/[\gamma(\gamma-1)]$ occurs on a time-scale $t_\mathrm{rel}(R) = \beta/\Omega(R)$ as
\begin{align}
    \Lambda=\left(\pd{\delta e}{t}\right)_\mathrm{rel} = - \frac{\delta e}{t_\mathrm{rel}},
    \label{eq:beta_cool}
\end{align}
where $\delta e$ is the thermal energy perturbation. Here $\beta$ is a spatially constant parameter (dimensionless cooling time), which we vary between models. We implement this term in Athena++ by retrieving the internal energy density from the total energy density, which is then changed each time-step using the exact solution to equation (\ref{eq:beta_cool})
\begin{align}
    \delta e(R,\phi,t+\Delta t) = \exp\left( - \Delta t/ t_\mathrm{rel} \right) \delta e(R,\phi,t)
    \label{eq:sol_beta_cool}
\end{align}
and added back to the kinetic term to update the total energy density. The implementation of this additional effect is verified by comparison with the linear solver.

The prescription (\ref{eq:beta_cool}) includes two important limits: for $\beta \rightarrow \infty$ the equation of state becomes adiabatic (i.e. individual fluid elements conserve their entropy in the absence of shocks), e.g. as in our fiducial model; for $\beta \rightarrow 0$ one recovers the \textit{locally} isothermal limit such that $\cs = \csi(R)$. Depending on the thermodynamic model, the effective sound speed varies, as a function of $\beta$ \citep{Miranda2020I}.

In practice, we set the cooling parameter $\beta = 10^2$ to achieve a situation close to an adiabatic limit. We tested and confirmed that these runs give virtually identical results to even greater $\beta = 10^3$, indicating convergence towards the adiabatic limit, in line with the findings of \citet{Miranda2020I}.


\bsp	
\label{lastpage}
\end{document}